# Plasmonic and Metamaterial Structures as Electromagnetic Absorbers

Yanxia Cui[1,2], Yingran He[1], Yi Jin[1], Fei Ding[1], Liu Yang[1], Yuqian Ye[3], Shoumin Zhong[1], Yinyue Lin[2], Sailing He[1,*]

[1] State Key Laboratory of Modern Optical Instrumentation, Centre for Optical and Electromagnetic Research, Zhejiang University, Hangzhou 310058, China
[2] Key Lab of Advanced Transducers and Intelligent Control System, Ministry of Education and Shanxi Province, College of Physics and Optoelectronics, Taiyuan University of Technology, Taiyuan, 030024, China
[3] Department of Physics, Hangzhou Normal University, Hangzhou 310012, China
Corresponding author: e-mail sailing@kth.se

Abstract: Electromagnetic absorbers have drawn increasing attention in many areas. A series of plasmonic and metamaterial structures can work as efficient narrow band absorbers due to the excitation of plasmonic or photonic resonances, providing a great potential for applications in designing selective thermal emitters, bio-sensing, etc. In other applications such as solar energy harvesting and photonic detection, the bandwidth of light absorbers is required to be quite broad. Under such a background, a variety of mechanisms of broadband/multiband absorption have been proposed, such as mixing multiple resonances together, exciting phase resonances, slowing down light by anisotropic metamaterials, employing high loss materials and so on.

## 1. Introduction

Electromagnetic (EM) wave absorbers are devices in which the incident radiation at the operating wavelengths can be efficiently absorbed, and then transformed into ohmic heat or other forms of energy. Thereby, neither transmission nor reflection is produced when a wave passes through a perfect absorber. Traditional absorbing layers are made from materials with high intrinsic losses, but EM absorbers that concern us are fashioned mainly from noble metals, such as gold, silver, or copper.

Mechanisms to construct EM absorbers are versatile. Early research on EM absorbers dates back to 1902 when Wood observed the anomalous dips in the reflection spectra of metallic gratings under illumination of a white light source [1]. Such kinds of metallic absorbers are comprised of arrays of metallic elements with periodicity on the order of the incident wavelength at optical frequencies [2-5]. They absorb light mainly due to the excitation of surface plasmon polaritons (or SPPs, for short), which are collective oscillations of free electrons supported at the interface between the metal and dielectric [6,7]. Due to the excitation of SPPs, the incident energy can be tightly trapped in the near field region so that the incident light can be fully absorbed at certain wavelengths by the metal/dielectric structures with specially designed geometries. There are various types of configurations being used as EM absorbers, such as lamellar gratings [8], convex grooves [9], spherical voids [10], and hole arrays [11,12]. These absorbers are made of noble metals, and associated with plasmonics that contains interesting physical phenomena associated with planar or localized SPPs [13,14].

Metamaterials are artificial assemblies of structured elements of subwavelength size (i.e., much smaller than the wavelength of the incident waves) [15]. They are often described as effective materials [16] and such a description is very convenient for an intuitive understanding of the phenomenon. The effective permittivity and permeability can be designated from zero to infinity, and as a result, various unique properties that are not available in nature can be finally achieved. Metamaterial devices inherently bear absorption loss on account of the imaginary parts of optical indices, degrading the performance in most applications such as super-resolution imaging, sensing, cloaking, etc. However, the absorption loss becomes advantageous for EM absorbers, and perfect absorption can be achieved by engineering the geometry of the structured elements [17]. In 2008, a perfect EM absorber based on metamaterials was first reported at the microwave regime [18]. Since then, EM absorbers have been developed further into the deep subwavelength scale and can still suck up the incident radiation with near-unity absorptivities [19-30]. At optical frequencies, metamaterials formed of metallic subwavelength structures are sometimes called plasmonic metamaterials when the excitation of localized SPP modes is essential in the process of light absorption [31,32]. Benefiting from advanced nanofabrication technologies, metamaterial-based EM absorbers are experimentally possible with enhanced miniaturization, increased adaptability, and increased effectiveness.

EM absorbers based on metallic structures have distinct advantages in comparison with their traditional counterparts. At different frequency ranges, they have different potential applications. As is well known, absorbing devices are critical components in solar energy harvesting systems, which have been investigated over the past half-century. So far, different types of solar absorbing coatings have been invented, including intrinsic, multilayer absorbers, textured low-bandgap semiconductors, metal-dielectric composite coatings, etc. [33-37]. Today, solar energy harvesting devices, embedded with metallic nanostructures, are promising candidates for harvesting energy with improved efficiency, wider solar spectrum, and lower material consumption, and are attracting widespread interest [38-42]. At microwave frequencies, absorbing devices, initially comprised of a metal ground plane, a dielectric spacer and a top resistive sheet [43], were largely applied in the military to provide concealment. However in modern designs of microwave absorbers, the metamaterial-based EM absorbers without the resistive sheets have equal or better performances than those with the resistive sheet [18]. According to Kirchhoff's law, a perfect emitter is equivalent to a perfect absorber. Thus, selective thermal emitters can be designed based on exactly the same principle as that of selective absorbers at infrared frequencies. Typically, semiconductor thin films [44], semiconductor gratings [45] and tungsten photonic crystals [46,47] are used to compose selective thermal emitters. Recently, a series of thermal emitters have been proposed based on subwavelength metallic structures because this method introduces more freedom to tailoring the emission spectrum [48-50]. In addition, by dealing with the fine spectra features of EM absorbers, researchers have successfully put forward many different biosensor configurations with greatly enhanced sensitivity to refractive index variation [51,52]. Moreover, recent studies have examined how to obtain EM modulators based on absorption [53-55].

This review is mainly to describe the principle of different types of narrow band EM absorbers as well as the various approaches to achieve broadband/multiband absorbers. It is mentioned that the scope of our work is different from a recent article [17] that mainly reviewed the researches on EM absorbers based on metamaterials. Instead, we include many mechanisms of EM absorption based on metallic structures, rather than only the metamaterial-based schemes. In Section 2, we start with the simplest EM absorber comprised of planar metal/dielectric stacks, followed with plasmonic absorbers composed of metallic gratings with the metamaterial based EM absorbers. Section 3 demonstrates how to improve the performance of the absorption band. Different proposals have been overviewed, including mixing multiple resonances together, exciting phase resonances based on lattice scattering, utilizing metamaterials to slow down the group velocity of light, employing high loss materials, etc. Then, Section 4 presents some examples of applications of EM absorbers with either narrow or broad bandwidths. Finally, this review ends in Section 5 with a summary of this work as well as our outlook on the future challenges of research on EM absorbers based on metallic structures.

## 2. EM absorbers with narrow bandwidth

Early research on EM absorbers focused on those with narrow absorption bands because their designs are relatively simple and the principles of absorption are inseparable from some fundamental phenomena of interference optics, plasmonics or metamaterials. Subsection 2.1 presents the first branch of narrow band absorbers in a straight format. Subsection 2.2 describes EM absorbers based on metallic gratings, which have the oldest history. The corresponding principles divide into two categories: planar SPPs and localized SPPs. Nowadays, the most popular absorbers are metamaterial absorbers that have experienced an explosive growth of interest for the past few years. In subsection 2.3, we will focus on introducing a few typical designs of metamaterial absorbers and the corresponding physical origins of full absorption. Section 2.3 ends with the procedure of making the narrow band absorber ultrathin.

### 2.1 Planar metal/dielectric stacks as absorbers

The simplest EM absorber is constructed based on a one-dimensional (1D) planar stack that requires very little nanofabrication. In 2006, Lee and Zhang [56] explored a simple three-layer structure, comprised of a 2 $\mu$m thick silica layer sandwiched by two silver films, as shown in Fig. 1a. By this structure, they obtained a close-to-unity thermal emitter at the infrared frequency. The dielectric medium located in the middle functions as an optical cavity of finite length, and the top and bottom metallic films sandwiching the cavity function as two non-perfect mirrors. The bottom metal film should be thick enough to block light transmission, while the top metal layer is required to be ultrathin (15 nm, close to the radiation penetration depth) to enable transmission of the incident wave into the cavity and further generate resonance.

The geometry of the cavity appears asymmetrical, but this structure is in fact a symmetric FP cavity because both its top and bottom boundaries are adjacent to metal films. Its resonant condition is approximately determined by the following formula considering the metal films as perfect mirrors:

$$2nk_0 d = 2m\pi, \quad m = 1, 2, \ldots \qquad (1)$$

where $n$ and $d$ are the refractive index and thickness of the middle dielectric, respectively, and $k_0$ is the propagating constant of light in vacuum. Here, the light path for one round trip within the cavity is an integral multiple of $2\pi$ since perfect mirrors, which are perfect electric conductors (PECs), force the electric field along the tangential direction at the PEC interfaces to be zero. In practice, the two films sandwiching the silica layer are made of real metals with finite optical conductivity, yielding light penetration into them. Besides, the top metal film is very thin, allowing the interaction between light within and outside the cavity. Both of these practical conditions introduce additional optical phase accumulations; accordingly, the resonant condition of Eq. (1) requires modifications.

Given that the structure is stratified, one can use some mature film fabrication technologies, such as magnetron sputtering and e-beam evaporation, to fabricate the required sample. A sample working at visible frequencies by e-beam evaporation was fabricated, and it composes of silver (30 nm)/silica (170 nm)/silver (300 nm) films on top of a glass substrate. The optical constants of noble metals used here are obtained from Palik's optical constant book [57], and the refractive index of silica is set to 1.52. This design works as a perfect optical absorber at $\lambda_0 = 674$ nm at normal incidence, within the considered wavelength range, as shown by the curve with circular symbols in Fig. 1b. In theory, the reflection/absorption spectra and field distribution can be straightforwardly calculated according to the transfer matrix method based on the Fresnel equation [58,59].

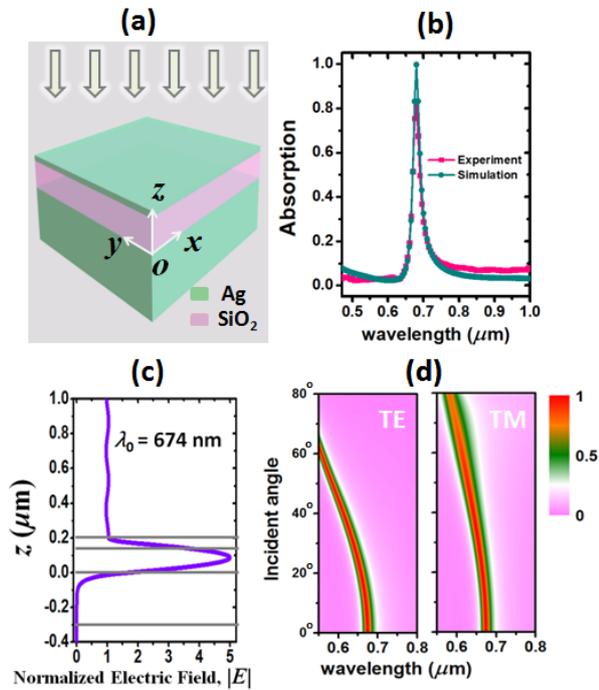

Figure 1 (a) Configuration of a narrow-band EM absorber based on planar metal/dielectric composite stacks. (b) Absorption efficiencies of a planar absorber composed of Ag/SiO$_2$/Ag on a glass substrate. (c) Field distribution at the absorption peak. (d) Absorption spectra via the incident angle for TE and TM polarized illuminations.

The absorption spectrum obtained from experimentally measured reflectivities is also plotted in Fig. 1b by the curve with rectangular symbols. As indicated, the simulation and experimental results agree with each other very well considering the peak envelope and peak position, although the experimental data of the absorption efficiency at the resonant peak is a bit lower than the simulated one due to the undesirable rough surface morphologies during fabrication. Fig. 1c shows the distribution of the normalized electric field ($|E|$) at different positions within the device. A cavity resonance with a very strong field localized between the two silver mirrors can be identified. According to Eq. (1), for this structure, the absorption peak at $\lambda_0 = 674$ nm is the 1$^{st}$ order resonance (i.e., $m = 1$). Within the incident space, the field has the amplitude close to that of the incident light ($|E|\sim 1$)), much smaller compared to that within the cavity. In addition, it suffers negligible oscillation, illustrating that the reflection from the device is almost cancelled. When Eq. (1) is fulfilled with $m$ larger than 1, FP resonances of higher orders can be excited, but with large frequency gaps between neighbouring absorption peaks. At normal incidence, because of the flat property of the geometry, this narrow band absorber performs the same for TE polarization (with the electric field perpendicular to the wave propagation plane) and TM polarization (with the magnetic field perpendicular to the wave propagation plane). When the incident angle becomes large, the absorption peak displays a notable wavelength shift but still owns near perfect absorption efficiency for both polarizations as shown in Fig. 1d.

There is another type of planar absorbers by putting a dielectric thin film/structure on a noble metal substrate [60]. In Ref. [60], the thickness of the ultrathin highly absorbing dielectric is far less than a quarter of the wavelength of light propagating within the dielectric media. The reason of light absorption was ascribed to the non-trivial interface phase shifts produced by the large optical attenuation within the highly absorbing dielectrics. They called it an asymmetric FP cavity as only the bottom of the cavity is adjacent to metal. In that work, examples of such kind of absorbers working at visible frequencies have been demonstrated by coating a nanometer film of germanium on top of gold substrate. By tuning the germanium film thickness from 7 to 25 nm, the colour of the sample changes from light pink to light blue.

The above discussions utilize the interference theory to explain the absorption of light. However, the essential reason behind these problems is the excitation of planar SPPs. SPPs are collective oscillations of free electrons supported at the interface between a metal and dielectric. Thereinto, planar SPPs are homogeneous solutions of Maxwell's equations for multilayer systems. Historically, planar SPPs supported at a single metal/dielectric interface were the first and simplest SPP modes. It satisfies the following dispersion equation [6,13]

$$\beta = k_0 \sqrt{\frac{\varepsilon_1 \varepsilon_2}{\varepsilon_1 + \varepsilon_2}} \qquad (2)$$

where $\beta$ is the propagating constant of the mode along the interface direction, $k_0$ is the wave vector in vacuum, and $\varepsilon_1$ and $\varepsilon_2$ are the relative permittivities of the metal and the dielectric, respectively. It is found that its maximum magnetic field intensity is located at the interface, and the field decays exponentially into both neighboring media. Hence, it belongs to a kind of surface wave. It only has mode confinement along the direction perpendicular to the interface; along the interface direction, light has typical propagating properties. Therefore, it is termed as planar SPPs (or propagating SPPs).

For SPPs propagating at a flat metal/air interface, because of the existance of momentum mismatch ($\beta$ is larger than the propagating constant in air), it is impossible to excite the propagating SPPs directly. Techniques of attenuated total internal reflection based on bulky prisms is one way to compensate for the phase difference [6]. It is also feasible to excite planar SPPs by etching the surface into the periodical format as discussed in 2.2.2. Another method is by modifying the metal/air one-interface system into multilayer systems with at least two interfaces as mentioned in Refs. [56] and [61]. By this approach, the dispersion band of these planar SPPs can be moved to have direct interaction with the incident light (i.e. the momentum match is established). It is emphasized that such kind of planar metal/dielectric stacks are very easy to fabricate and have already found considerable applications in practice as shown in Section 4.

**2.2 Reflective metallic gratings as absorbers**

Reflective metallic gratings have the fascinating property of efficiently absorbing the incident light. So far, it is generally accepted that the absorption of light by reflective metallic gratings approaches unity as a result of the excitation of SPPs. SPPs can be classified into two different types: planar SPPs with only 1D mode confinement as mentioned in Section 2.1 and localized SPPs with at least 2D mode confinement. This section begins with a brief history of research on reflective metallic gratings in subsection 2.2.1. It is followed by an introduction of the mechanism of total absorption by reflective metallic gratings due to planar SPPs (subsection 2.2.2). Afterwards, the physical origin of total absorption produced by the excitation of localized SPPs along with a recent progress on this topic will be presented in subsection 2.2.3.

**2.2.1 History of research on reflective metallic gratings**

The history of reflective metallic grating begins with the discovery of grating anomalies by Wood in 1902 [1], which have been thoroughly investigated in the past one hundred years. A good historical introduction of studies related to Wood's anomaly can be referred to Ref. [62] written by D. Maystre. In the following, we only present some very important historical developments.

In 1907, Rayleigh explained Wood's anomaly according to the grating formula [63], and conjected that the reflection of light passes off at sharply defined wavelengths. And this type of reflection dips is purely dependent on the grating period. In 1941, Fano made a breakthrough on explaining Wood's anomaly and addressed that, in addition to Rayleigh's conjecture, there exists a forced resonance supported by the specific metallic grating which causes the reflection valleys to broaden at wavelengths other than those predicted according to the grating formula [64]. In 1965, by using numerical tools, Hessel and Oliner achieved the same conclusions as Fano [65]. In 1973, thanks to the invention of holographic technology, an experimental study on metallic gratings with accurately controlled geometries became possible [66]. Later, the first agreement between the numerical and experimental results of diffraction efficiency of a specific metallic grating was achieved [67].

Note that it was Maystre who first developed the rigorous vector theory of gratings in 1972 [68]. This theory allows the study of Wood's anomaly based on the wavelength dependent permittivites of metallic materials, recognizing that the perfect conductor model is not valid at optical frequency. Nowadays, numerical methods including the finite element method (FEM) [69], finite difference time domain method (FDTD) [70], finite integration technique (FIT) [71], and rigorous coupled wave analysis method (RCWA) [72] have become powerful enough to quantify the electromagnetic response from arbitrary objects. More importantly, they have been packaged into commercial softwares, such as Comsol Multiphysics, Lumerical Solutions, CST, Rsoft, most of which have very good user interfaces. Such facilities enable scientists to step directly into discovering the physics of the problem and thus speed up the research progress. The aforementioned numerical tools have been popularly used to calculate the diffraction efficiencies as well as the field distributions of metallic gratings.

Despite the fact that Wood's anomalies show obvious dips in reflection spectra, reflection at these dips were not fully cancelled in the early experimental study. Until 1976, the phenomenon of total absorption of TM-polarized light was firstly demonstrated by Hutley and Maystre [73,74]. In detail, they displayed a minimum reflectance of 0.3% at an angle of incidence of 6.6º for a gold sinusoidal grating with a period of 555 nm and a depth of 37 nm, illuminated by a TM polarized light at $\lambda_0$ = 647 nm. At that time, such a finding was very surprising as very gentle modulation in the surface of a gold mirror leads the reflectance to fall dramatically to below 1%. The interpretation of the absorption phenomena is due to the resonant excitation of planar SPPs with only 1D mode confinement. One year later, total absorption of TE-polarized light was achieved by coating the shallow metallic grating with a thin dielectric layer [75,76]. This curious absorption is not due to the excitation of planar SPPs but a waveguide mode propagating in the dielectric film. EM absorbers based on propagating surface waves are more like a result of diffraction behaviours, which are very selective to the incident angle at a given investigated wavelength. Note that selective absorption can be used in metrology as well as making biosensors. About a decade ago, studies of metallic grating entered a new era when localized SPPs became capable of trapping light [4,8,77,78]. Such absorbers perform with angular insensitivity because they have at least 2D mode confinement, which is crucial in the application of solar energy harvesting.

**2.2.2 Total absorption of metallic grating due to planar SPPs**

One ways to solve the problem of momentum mismatch for the excitation of planar SPPs is by using a diffraction grating, which changes the in-plane wave vector of the incident light field [6]. With the addition of the grating wave vector, phase matching takes place when

$$\beta = k_0 \sin\theta \pm m\frac{2\pi}{a}, \quad m = 1, 2, \ldots \quad (3)$$

is fulfilled, where $\theta$ is the incident angle and $a$ is the grating period. It is worth mentioning that with the grating pattern, the envelope of SPPs is retained but the propagating constant of the exact SPPs supported on the metallic grating has been slightly changed from that at the flat interface. In other words, the propagating mode excited on the surface of a shallow metallic grating is in fact a distortion from the SPPs supported at a single interface between the metal and dielectric.

Initial investigations of planar SPPs were done on the surface of sinusoidal shaped metallic gratings because the corresponding theoretical treatment was easier [73,74]. To excite planar SPPs, the metallic grating can also be adapted into other shaped profiles. Here, we show an example of rectangular shaped grooves (period $a = 700$ nm, width $w = 165$ nm, and height $h = 34$ nm) on a 300 nm thick flat gold substrate (Fig. 2a). Its absorption spectra at normal incidence for TM-polarized light is plotted in Fig. 2b, which indicates that at $\lambda_0 = 750$ nm the metallic grating can completely transform the incident light into planar SPPs along the grating vector direction. Its magnetic field distribution at the absorption peak is shown in Fig. 2c, in which it is seen that the field has been successfully bounded to the gold surface. The field in metal becomes evanescent very quickly but the field in the air can extend much farther away from the gold surface. According to Eq. (3), the propagating constant of the excited mode in Fig. 2c is $2\pi/a$, which is a bit larger than that supported on a flat metal surface according to Eq. (2) ($\lambda_{sp} \approx 714$ nm). The interpretation is that etched grooves have prolonged the optical path of light, reducing the propagating constant. Bear in mind that based on Eq. (3), its phenomenon of total absorption results in a very strong angular selectivity (Fig. 2d).

An alternative approach to excite planar SPPs is to design a dielectric grating on top of a flat metallic substrate [79], which can save the fabrication cost in comparison with the aforementioned device with grooves etched in metals. An example can be found in Fig. 2e, in which a silica grating with period $a = 700$ nm, width $w = 156$ nm, and height $h = 250$ nm is placed on a 300 nm thick, flat gold substrate. For normal incident TM-polarized light at 750 nm, the reflection becomes zero and perfect absorption takes place as shown in Fig. 2f. It is found in Fig. 2g that the field distribution at the absorption peak maintains the envelope of planar SPPs. Note that the bandwidth of the absorption peak in Fig. 2f is narrower than that in Fig. 2b. This might be attributed to the SPPs propagating path. On the metallic grating, the path is lengthened due to the existence of metallic grooves [80], and thus the propagating loss of planar SPPs in Fig. 2c is larger than that for the dielectric grating in Fig. 2g, of which the propagating path is flat. Smaller propagating loss means a higher quality factor of the resonance, and thus a narrower absorption peak. Absorbers comprised of dielectric grating loaded on a flat metal film show the property of angular sensitivity as well (see Fig. 2h).

As mentioned in Section 2.2.1, total absorption of TE-polarized light can be achieved by coating the metallic grating with a thin dielectric layer due to the excitation of a waveguide mode propagating in the dielectric film [75,76]. A metallic grating, the same as mentioned in Fig. 2a, covered by a thin dielectric film made of silica with a height of 176 nm (from the top surface of the silica film to the bottom of the groove is 210 nm), is investigated. Fig. 2i-2l show its configuration, absorption spectra at normal incidence, field distribution at the absorption peak ($|E|$), and the angular spectra, respectively. At normal incidence with a wavelength of 750 nm at TE-polarization, perfect absorption indeed takes place. Fig. 2k shows that, at resonance, a typical waveguide mode is excited in the dielectric thin film, which follows a zigzag path with a lossy metallic reflection on one side and a total reflection on the other side. Performance of such an absorber is again sensitive to the incident angle (see Fig. 2l).

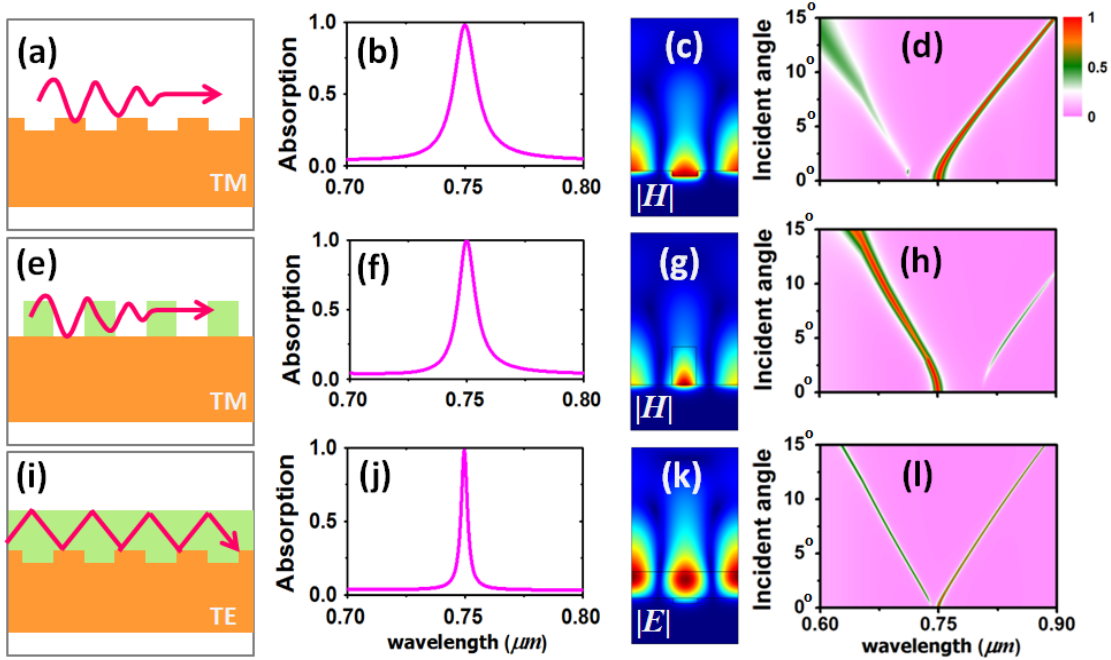

Figure 2 Configuration, absorption spectra at normal incidence, field distribution at the absorption peak, and the angular spectra for (a-d) the shallow metallic grating, (e-h) the dielectric grating loaded on a flat metal substrate, and (i-l) the metallic grating coated with a thin dielectric film. The first two types of absorbers work at TM polarization, and the third one works at TE polarization.

A modified configuration is obtained by introducing a dielectric grating on top of the dielectric film that is placed on a metal plate [81]. Recall that a good spectrum feature requires the refractive index of the dielectric film to be larger than that of the dielectric grating. Such a modified structure supports not only the planar SPPs at TM polarization similar to that discussed in Fig. 2g, but also higher order waveguide modes for both TM and TE polarizations when the dielectric film is thick enough, since the excitation of waveguide modes requires the dielectric film thickness to be larger than the corresponding cutoff thicknesses. Finally, rich spectral features with multiple absorption peaks can be produced by this structure.

To summarize, total absorption of light by the gratings discussed in this section could be attributed to the propagating modes of either planar SPPs or waveguide modes. The grating embedded at the interface of the absorber is responsible for the momentum matching, which induces apparent angular selectivity of the performance. It is worth mentioning that if the metallic grating is extended into biperiodic grating (i.e. crossed grating), the absorption of unpolarized light at a certain incident angle can occur [82].

### 2.2.2 Total absorption due to localized SPPs

The main feature of localized SPPs is that once it is excited, the energy accumulates in a local isolated region, hence its name. The geometry of the individual absorbing elements, especially the part that confines light, determines the resonance, while the effect of the interaction between neighboring absorbing elements is weak. Rectangular shaped metallic grooves [5,8,83-91] are the most frequently investigated configurations of efficient absorbers based on the excitation of localized SPPs, as shown in Fig. 3a. This type of grooves is different from those depicted in Fig. 2 in that the groove depth is optically deeper in comparison with the effective wavelength of the SPP mode propagating inside the grooves.

In fact, a deep rectangular shaped groove can be regarded as a truncated system from the classical metal-insulator-metal (usually shortened to MIM) multilayer system. A MIM multilayer system comprises of a dielectric layer of a finite small thickness sandwiched between two metal layers. As discussed in the previous subsection, each single metal/dielectric interface can sustain a confined wave of planar SPPs according to Eq. (2). When the separation between adjacent interfaces is small enough for inducing interactions between two planar SPPs, it gives rise to coupled modes of symmetric or anti-symmetric profile with respect to the magnetic field distributed at the two interfaces. Such modes, propagating along the interface direction, belong to waveguide modes, with the field mainly confined inside the middle dielectric region. As a general rule, the symmetric waveguide mode takes responsibility of absorption in the metallic groove. It satisfies the dispersion relation of

$$\tanh(k_2 \frac{w}{2}) = -\frac{k_1 \varepsilon_2}{k_2 \varepsilon_1}, \quad \text{with } k_{1,2}^2 = \beta^2 - k_0^2 \varepsilon_{1,2} \quad (4)$$

where $\beta$ is the propagating constant of the symmetric mode along the interface direction, $k_0$ is the wave vector in vacuum, and $\varepsilon_1$ and $\varepsilon_2$ are the relative permittivities of the metal and the dielectric, respectively. This kind of SPP waves is always termed as gap SPPs, because the insulator can be regarded as a gap of a continuous metal film. For gaps with small widths, a simple analytical expression of the propagating constant of the symmetric mode is obtained by some approximations based on Eq. (4) [92,93]:

$$\beta = k_0 \sqrt{\varepsilon_2} \left(1 + \frac{\lambda_0}{\pi w \sqrt{-\varepsilon_1}} \sqrt{1 + \frac{\varepsilon_2}{-\varepsilon_1}}\right)^{1/2} \quad (5)$$

It is noted that Eq. (5) is not applicable when the gap width is too small [92]. We note that this symmetric waveguide mode has no cut-off width, and the effective refractive index $n_{\text{eff}}$ (i.e. $\beta/k_0$) keeps increasing when the waveguide becomes narrow.

For a single metallic groove, regarded as a truncated system from the MIM waveguide, its top end faces the incident space (air). At the plane of its top end, total internal reflection occurs because the effective refractive index of the propagating mode inside the groove is larger than unity. Its bottom end adheres to a blocking metal, regarded as a non-perfect mirror. When an ideal mirror is considered, the electric field in the tangential direction is equal to zero. In practice, metals with inherent loss allow light to penetrate into them by skin depth, which corresponds to an additional optical path length. Therefore, a closed cavity is formed with 2D mode confinement and thus a standing wave is generated inside the groove.

The resonant condition of a rectangular groove can be approximately written as

$$2n_{\text{eff}} k_0 d + \varphi_{\text{up}} + \varphi_{\text{low}} = (2m-1)\pi, \quad m = 1, 2, \ldots \quad (6)$$

where $n_{\text{eff}}$ is the aforementioned effective refractive index of the gap SPP mode obtained by solving Eq. (4), $d$ is the groove depth, and $\varphi_{\text{up}}$ and $\varphi_{\text{low}}$ are the additional phases generated due to the reflection at the opening and the penetration at the bottom, respectively. Different from a typical FP cavity, a $\pi$ phase shift is involved in the right side of Eq. (6) due to the introduction of the mirror plane. When $m = 1$, ignoring any additional phase ($\varphi_{up} = \varphi_{low} = 0$), the product of $n_{eff}$ and $d$ is a quarter-wavelength. It is worth noting that a metallic groove with a depth of $d$ corresponds to a metallic slit [94] with the same width and a depth of $2d$, which satisfies a resonant condition of

$$2n_{eff}k_0 d + \varphi_{up} = m\pi, \quad m = 1, 2, \ldots \quad (7)$$

By comparing Eqs. (6) and (7), we know that resonant modes of even orders existing in a metallic slit of depth $2d$ disappear in a metallic groove of depth $d$. That is again attributed to the existence of the mirror plane at the bottom of the groove.

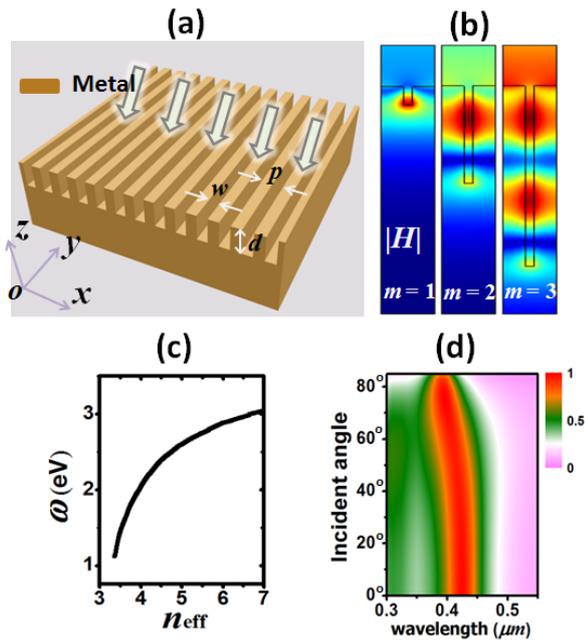

Figure 3 (a) Configuration of a metallic absorber with deep rectangular grooves. (b) Field distributions for the grating (with $w = 5$ nm and $p = 30$ nm) normal-illuminated at $\lambda_0 = 475$ nm. Left: $h = 15$ nm, Middle: $h = 80$ nm, Right: $h = 148$ nm. (c) The dispersion diagram ($\omega_0 \sim n_{eff}$) of an Ag/air (5 nm)/Ag waveguide. (d) Angular absorption spectra when $h = 15$ nm.

Ref. [5] reported that light can be strongly absorbed with an efficiency of 98% at $\lambda_0 = 475$ nm ($\omega_0 = 2.6$ eV) even when the silver grooves of period $p = 30$ nm are a few nanometers deep ($d = 15$ nm) and wide ($w = 5$ nm). It shows that, although the groove is geometrically shallow, the near unity absorption is due to the excitation of gap SPPs because the groove width is very small. The left subplot of Fig. 3b shows the magnetic field distribution at its absorption peak, from which one sees that the tiny groove confines the strong field inside of it. Fig. 3c shows the calculated dispersion diagram of $\omega_0 \sim n_{eff}$ for an Ag/air (5 nm)/Ag waveguide based on Eq. (4), which indicates that the effective refractive index of the gap SPP mode at the studied wavelength is quite large ($n_{eff} = 4.97$). Then, the calculated product of $n_{eff}$ and $d$ is 74.6 nm, which is close to a quarter (118.8 nm) of the incident wavelength (corresponding to the $1^{st}$ order FP resonance). The remaining difference lies on the additional optical path generated at the opening and bottom of the groove since the effect of light penetration at the bottom of the groove can be clearly seen from the left subplot of Fig. 3b. Consequently, the near unity absorption is, with no surprise, due to the excitation of gap SPPs. By tuning the groove depth, it excites higher orders of FP resonances. For example, when the groove depth is 80 or 148 nm, the absorption peak of $m = 2$ or 3, respectively, is excited but with an efficiency much lower than unity. The middle or right subplot of Fig. 3b shows the corresponding field distribution for the absorption peak of $m = 2$ or 3, respectively, indicating that light propagating in the resonant groove of the $m^{th}$ order indeed suffers an additional $2\pi$ phase shift per round trip compared with that of the $(m-1)^{th}$ order.

In addition, because the geometry of the groove plays the significant role of determining the resonance and the grating has a very short pitch, a flat SPP band is produced and the corresponding absorption peak is almost insensitive to the angle of incidence (see its angular absorption spectra when $d = 15$ nm in Fig. 3d). However, when the grating period is larger than half the incident wavelength, the propagating SPP band following the rule of Eq. (3) will cross the flat SPP band and thus form a compound SPP band which is dispersive around the crossing point [83,95]. Moreover, Rayleigh anomalies [96-98] can suppress the resonance of gap SPPs when the following condition is satisfied:

$$m\lambda = a(1 \pm \sin\theta), \quad m = 1, 2, \ldots \quad (8).$$

If the Rayleigh anomaly is excited, light will be reflected grazing the grating surface, and the field distribution pattern will resemble a distortion of the constructive interference pattern formed by two plane waves propagating along the grating surface. Such an anomaly produces a very sharp absorption peak with a bandwidth as small as 0.2 nm, which might be promising in the area of highly sensitive sensors.

The rectangular shaped groove is oriented along the vertical direction, while in recent years the horizontally oriented MIM waveguide cavity has become a widespread interest [31,51,99-114]. It consists of three layers: the bottom metallic substrate with sufficiently large thickness to block light, the center dielectric layer, and the very thin top metallic layer fashioned in strips in 2D geometry or variously shaped patches in 3D geometry. Such a structure was first theoretically analyzed in 2006 [99], and then the experimental investigation on its absorption property was demonstrated in 2009 [102]. Later, we adopted them to enhance the extraordinary transmission of light through a metallic nanoslit [103,115]. They can be called as "nano cavity antennas" considering their ability of harvesting electromagnetic waves like antennas at radio frequency. The lithography technique is the typical method to fabricate the patterned metallic structures, which is expensive and complicated. In 2012, Moreau *et al.* created a MIM cavity absorber by randomly distributed chemically synthesized silver nanocubes onto a nanoscale thick polymer spacer layer on a gold film, which makes the manufacture simple, rapid, inexpensive and easily scalable [114].

Here, we show a 2D case of gold/silica/gold multilayer system as shown in Fig. 4a with $p$ = 265 nm, $d$ = 20 nm, $w$ = 80 nm, and $h$ = 18 nm. This system can efficiently absorb light at 750 nm with an efficiency of about 95% (its calculated absorption spectra is shown in Fig. 4b). Fig. 4c shows its magnetic field distribution at the absorption peak, from which one can observe a representative 1$^{st}$ order gap SPP cavity mode sustained in a MIM waveguide of finite length. It should be noticed that the thickness of the top layer of the metallic strip is critical as it should be not only thick enough for mode confinement along the vertical direction but also thin enough for efficient coupling of light into the horizontal cavity. This resonance is almost dispersionless as shown in Fig. 4d. In Subsection 2.3, the absorbers based on MIM waveguide cavities will be mentioned again as such tiled cavities can be regarded as a thin layer of metamaterial [31,106,111].

In the above discussions, all the MIM waveguide cavities have a fixed width, which simplifies the problem. In reality, localized SPPs were first recognized in a very complicated system comprised of a chain of semicylinders placed upon a silver surface in 1996 [116]. Soon afterwards, Samble's group [4,77,78,117,118] focused their research on short pitch metallic gratings, known as zero order gratings or nondiffractive gratings. That practice was effective for clarifying the mechanism of localized SPPs, as its resonance band was tuned to be divergent from the propagating SPPs band. The treated profiles of grooves in their works are usually written in functions, such as the Gaussian function, sinusoidal function, and superposition of multiple terms of the Fourier series expansion. They were the first group which exactly ascribed the full absorption to the excitation of MIM waveguide modes [4]. Since the width of the groove varies at different depths, some approximation was done in the analysis. Over that period, experimental realization of full absorption by a reflective metallic surface due to localized SPPs was also achieved [8,83]. Samble's group has also done a series of investigations on narrow ridged short pitch gratings [119-121], which support localized SPP modes with a symmetric charge distribution on either side of the grating ridges at angles away from normal incidence where strong resonant absorption occurs. Reflective metallic grating composed of V-shaped grooves whose width monotonously decreases with the increase of depth has also been investigated thoroughly [97,122-125]. It was found that V-shaped grooves can not only fully absorb light, but also focus light into their sharp corners at the bottom. Such extreme field localization can greatly enhance the electric field.

A 2D MIM waveguide cavity with its geometry invariant along the direction perpendicular to the incident plane only becomes resonant at TM-polarized incidence. For TE-polarized light, resonant absorption can occur due to the excitation of cavity modes in the same structure. It has been shown that by elaborately choosing the geometrical parameters of MIM waveguides, one can overcome the property of polarization dependance in 2D space through the excitation of localized SPPs for TM-polarized light and cavity modes for TE-polarized light simultaneously [126-128]. However, in order to completely eliminate its dependance on polarization in 3D space, the 1D array of elements should be extended into a 2D array. For the metallic grooves, there are two possibilities envisaged: one is with isolated metal blocks and the other is with continuous metal but isolated air holes. It is confirmed that the structure with isolated metal blocks is the evolution scheme from the 1D array of grooves [129,130]. However, the 2D array of metallic holes suffers the cut-off condition on hole sizes because cavity modes are excited inside the holes. For the horizontal MIM waveguide absorber as shown in Fig. 4a, a straightforward polarization-insensitive proposal is to fashion the top layer into a 2D array of rectangular patches [31,108,110,113]. Studies show that such a kind of MIM waveguide mode can still survive when the patches are in circular [51], cross [29,131], or even complicated dendritic shapes [132]. There are also a few studies dealing with the cases when the top thin metal layer is perforated with holes, which can also result in full absorption of light [49,133]. The 2D array of V-shaped grooves has also been investigated and shows properties similar to their 1D counterparts [125].

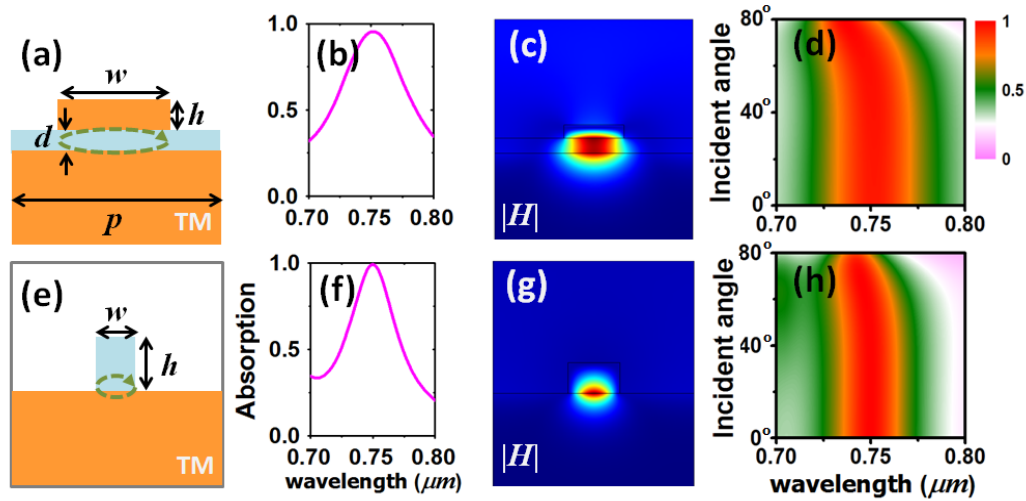

Figure 4 Configuration, absorption spectra at normal incidence, field distribution at the absorption peak, and the angular spectra for (a-d) the horizontal MIM waveguide cavity and (e-h) the dielectric grating of high refractive index loaded on a flat metal substrate.

It is interesting to find that the SPP wave excited on a flat metal/dielectric interface can also get concentrated in a local region under the condition that the SPP wave meets reflective boundaries along the propagating direction. By increasing the refractive index of the dielectric in the configuration of Fig. 2e, we can succeed in this goal. Here, we load a silicon grating on top of a flat gold surface with $p$ = 265 nm, $w$ = 69 nm, and $h$ = 44 nm. The refractive index of silicon is obtained from Palik's book [57]. Fig. 2f shows its absorption spectra at normal incidence and it indicates that the structure can fully consume light at 750 nm. By checking its field distribution at the absorption peak (Fig. 2g), one sees that the field is localized exactly at the gold/silicon interface. It indicates that the difference between the effective refractive indices of the SPP wave at the gold/silicon interface and that at the gold/air interface is large enough to give total reflection of the SPP wave excited on the gold/silicon interface. In this situation, a horizontal metal/dielectric waveguide cavity of finite length is formed, and thus 2D mode confinement occurs. This absorption peak shows negligible angle dependance as well (Fig. 4h).

Next, we will simply discuss the buried dielectric cavities in metal which can also act as efficient absorbers. Teperik *et al.* has made an effort in achieving omnidirectional full absorption based on the excitation of so called void plasmons [134] on a flat metal surface. A layer of close-packed spherical voids is embedded beneath the metal/air interface by a distance in the skin-depth range [135,136]. However, void plasmons can hardly be defined as a type of surface wave. That is because the field, rather than being concentrated around the metal surface, is localized in the center of the void cavity [137] as what happens in the case of photonic cavity resonances [138]. Another piece of evidence is that the diameter of the spherical voids is larger than half the incident wavelength, which is not below the diffraction limit [134]. Bonad and Popov examined the response of a 1D array of dielectric cylinder cavities buried in gold, and found full absorption at TE-polarized light [139]. At resonance, its electric field profile in the 2D cylindrical cavity is very similar to that in the 3D spherical cavity [137]. One may choose to bury a closed rectangular-shaped dielectric cavity (i.e. a MIM waveguide of finite length) in metal; then full absorption of TM-polarized light can be achieved due to the excitation of localized SPPs in the dielectric cavity via a tunneling effect through the top-coated metal layer of skin-depth thickness. If a narrow channel of width approximately 5 nm is opened in the top-coated metal layer to connect the incident space with the dielectric cavity, an interesting hot spot effect can be excited within the channel which can also contribute to full absorption in the near-infrared frequency range [126]. We mention that it is even possible to use transmissive metallic slit arrays to achieve anomalous full absorption of light by the assistance of dielectric slabs [140,141].

In summary, different forms of reflective gratings which can produce total absorption of light due to the excitation localized SPPs or photonic modes have been reviewed. At TM polarization, cavities with 2D confinement are generally demanded for exciting localized standing waves, which are responsible for the dispersionless absorption band. The most frequently studied cavities are MIM waveguides of finite length. Angle-insensitive absorbers for TE-polarized light have also been refered in this section, for which the absorption mainly relies on the excitation of photonic cavity modes on the metallic geometries of a much larger dimension. It is worth mentioning that if the 1D array of absorbing elements for TM-polarized light is generalized to the model of a 2D array, polarization-independent absorption in 3D space can be achieved. At last, we have also introduced some recent interesting works based on the buried dielectric cavities in metal with tiny slit channels or transmissive metallic slit arrays.

**2.3 Metamaterial absorbers**

Metamaterials allow for explicit design of arbitrary effective permeability and/or permittivity. Since the first metamaterial-based absorber was invented in 2008, they have been intensively investigated from microwave to optical frequencies. A recent progress report has reviewed how to design such absorbers, the theory underlying their operations along with future challenges [17]. The first so-called perfect metamaterial absorber consists of a top metallic layer with electric ring resonators (ERRs), a dielectric slab and a bottom metallic layer with cutting wires. As such, polarized incident light with the *E* field along the cutting wire direction can be perfectly absorbed.

Later, researchers introduced four-fold rotational symmetry into the ERR-based absorbers to eliminate the dependence of absorption on the incident polarization [19,142-144]. In practice, absorbers composed of ERRs are not superior because both layers of metals require the etching process. Another type of metamaterial absorbers has a very simplified form, of which one of the metal layers does not have any pattern. For example, it can be formed by sandwiching an array of metallic patches of arbitrary shape and a continuous ground plate together with a dielectric slab [31]. In fact, they are the MIM cavities presented in subsection 2.2.2. There, they are considered plasmonic absorbers due to the excitation of localized SPP modes. Here, because they can be described as an equivalent layer with effective constitutive parameters of unusual values, they are named plasmonic metamaterial absorbers [29,111,132,145-149]. Note that the optically deep metallic grooves have also been treated as metamaterial absorbers in many works

[22,111,130,150,151]. In addition, some reports also use 1D metal/dielectric stacks [21], 2D metallic nanowires [52,152] or 3D metallic nanoparticles [32,153] to form metamaterial absorbers. Recently, metamaterials whose effective $\varepsilon$ or $\mu$ is near zero have also been applied in forming absorbers [154-157], especially ultrathin absorbers [156]. Furthermore, other groups employ spatial dependent $\varepsilon$ to design frequency selective absorbers [158] or optical black holes [159]. In this section, we include three specific examples of metamaterial absorbers based on MIM cavities, metallic nanowire arrays and metamaterials whose effective $\mu$ is near zero.

### 2.3.1 Plasmonic metamaterial absorbers composed of MIM cavities

Fig. 5a shows a typical plasmonic metamaterial absorber based on MIM cavities, similar to that shown in Fig. 4a but extended into 3D space. The geometrical parameters are set according to Ref. [31]. Due to the 4-fold symmetry of the geometry, this absorber is not polarization sensitive. At normal incidence, the absorption spectra for TE (with $E \perp S_{xz}$) and TM (with $H \perp S_{xz}$) polarizations are identical as shown by the thick curve in Fig. 5b. In addition, its high absorption performance is almost independent of the incident angle as plotted in Fig. 5c.

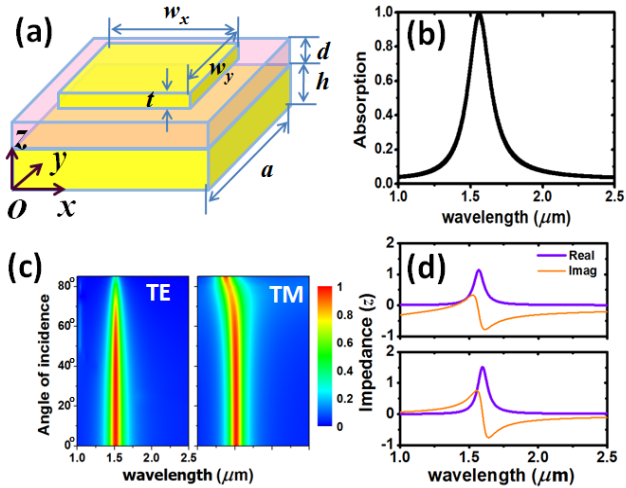

Figure 5 (a) Configuration of the plasmonic metamaterial absorber consisting of an array of MIM cavities. (b) Absorption spectrum of the absorber at normal incidence. (c) Angular absorption spectra at TE- and TM- polarization. (d) The retrieved impedances from the complex S parameters (top panel) and the critical coupling theory (bottom panel).

Its physical interpretation of absorption was addressed based on the effective medium theory [16], which assumes the absorber to be a single homogenous metamaterial slab, of which the effective constitutive parameters can be retrieved based on the complex $S$ parameters. It is noted that the retrieved effective parameters from a 2D planar surface with only one cell, would not hold for the corresponding 3D structures with multiple cells [16].

The top panel in Fig. 5d displays the retrieved impedances (normalized to the impedance of the free space) by treating the absorber as a single effective slab with a thickness of 100 nm. It indicates that at the resonant wavelength, the impedance of the slab with effective parameters approximately matches that of the free space. Because the complex $S$ parameters are not fully defined when transmittance vanishes, the effective medium theory does not work for metamaterial absorbers using a thick ground plate. Additionally, the thickness of the effective metamaterial slab is not uniquely defined, rendering the effective parameters arbitrary.

To avoid the aforementioned problems, Wu et al. explained the phenomenon of perfect absorption in terms of the critical coupling theory [106]. The metamaterial surface itself is regarded as a leaky resonator. Based on a single-resonance model, the impedance of the metamaterial surface is given by

$$z = \frac{\omega_{ie}}{-i(\omega-\omega_0)+\omega_{io}} \quad (9)$$

where $z$ is a function of the resonant frequency ($\omega_0$), the radiative damping/coupling rate ($\omega_{ie}$), and the resistive damping rate ($\omega_{io}$) of the resonator. According to Eq. (9), it is seen that in order to match the free space impedance, the radiative damping rate must be identical to the resistive damping rate. In other words, critical coupling occurs when the eigenmode of the resonator has equal resistive and radiative losses. Consequently, for the situation of perfect absorption, the incoming field excites the eigenmode of the resonator through the radiative coupling, and the incoming energy completely transforms into resistive loss. A unique feature of this method is its high computational efficiency because only a single eigenvalue simulation is required to determine the impedance as well as the absorption spectra by the formula of

$$A(\omega) = 1 - \left|\frac{z(\omega)-1}{z(\omega)+1}\right|^2 \quad (10)$$

Consider a corresponding two-dimensional (2D) system that completely blocks the transmission of light. By the critical coupling method, the impedance of the present 2D metamaterial absorber for TM polarization is obtained as shown on the bottom panel in Fig. 5d. It is seen that the results obtained by the critical coupling method are very close to the results calculated based on the complex $S$ parameters as shown on the top panel in Fig. 5d. At this moment, the effective parameters are no longer arbitrary. The eigenvalue simulation can also render the field distribution of the eigenmode at the resonant frequency ($\omega_0$), which is exactly the mode profile excited by an incident field. It provides evidence of localized magnetic resonance, for which the surface currents excited in the two metal layers are anti-parallel.

Dealing with the same problem, Koechlin et al. put forward the equivalent metamaterial treatment from a different angle of view [111]. Inspired by the work of Shen et al. [160], they describe periodic arrays of MIM cavities as an equivalent metamaterial, consisting of a single homogeneous layer deposited onto a perfect conductor. The effective index and effective thickness of this layer can be derived analytically from the parameters of the resonators. For MIM cavities with $w = 1.087$ μm, $d = 200$ nm, $a = 3.8$ μm, and the dielectric of germanium, they obtained the equivalent metamaterial with an equivalent index of $\bar{n} + \bar{k}i = 33.76 + 0.508i$ and equivalent thickness of $\bar{t} = 74$ nm. Then, the reflection coefficient is calculated according to the multi-beam interference theory

$$A = 1 - \left( \frac{r_{12} - e^{4i\pi(\bar{n}+i\bar{k})\bar{t}/\lambda}}{1 - r_{12}e^{4i\pi(\bar{n}+i\bar{k})\bar{t}/\lambda}} \right)^2 \quad (11)$$

where $r_{12} = (1 - \bar{n} - i\bar{k})/(1 + \bar{n} + i\bar{k})$ is the classical Fresnel coefficient. The calculated absorption spectrum by this mean is found to be in good agreement with the directly calculated results.

There are also some other viewpoints for explaining the principle of this type of perfect metamaterial absorber. Chen pointed out that the metasurface must be considered as a separate layer and concluded that the multiple reflections determine the perfect light absorption [148]. Pu et al. proposed a methodology based on the circuit theory that perfect absorption can be realized by matching the retrieved impedance of the metallic surface structure to that of the so-called perfectly impedance matched sheet [145]. Most recently, Costa et al. have analytically described the absorption efficiency in a formula related to electrical and geometrical parameters by resorting to the transmission line model [147]. It has been shown that the choice of highly capacitive-coupled elements (e.g. a rectangular-shaped patch array) allows obtaining the largest possible bandwidth, whereas a highly frequency-selective design is achieved with low capacitive elements (e.g. a cross-shaped patch array).

### 2.3.2 Metamaterial absorbers comprised of finite cavities made of a metallic nanowire array

In 2009, Kabashin et al. utilized metamaterials comprised of metallic nanowires to enhance the sensitivity to refractive-index variations of the medium between the rods [52]. Besides the application of sensing, metallic nanowire metamaterials can also be applied to harvesting energy [157]. We have proposed an efficient metamaterial absorber formed by periodic cavities composed of gold nanowires embedded in the alumina host as shown in Fig. 6a [152]. Here, the periodic cavities are grounded by a thick gold film to block light transmission, In each cavity, there are 6 by 6 gold nanowires.

According to the effective medium theory, the present structure can be treated as an effective medium with the following permittivity components,

$$\varepsilon_x = \varepsilon_y = \varepsilon_d \frac{(1+f_m)\varepsilon_m + (1-f_m)\varepsilon_d}{(1-f_m)\varepsilon_m + (1+f_m)\varepsilon_d}$$
$$\varepsilon_z = f_m\varepsilon_m + (1-f_m)\varepsilon_d$$
(12)

where $f_m$ is the volume filling ratio of gold nanowires. The absorption spectra are shown in Fig. 6b, where the blue curve and the red curve correspond to the absorber structures modelled with the realistic nanowire structure and the effective medium, respectively. It is found that the two absorption spectra are in agreement with each other, both exhibiting perfect light absorption around frequency $f = 195.9$ THz. Moreover, it is polarization-independent (due to the 4-fold symmetry of the geometry) and robust to the incidence angle.

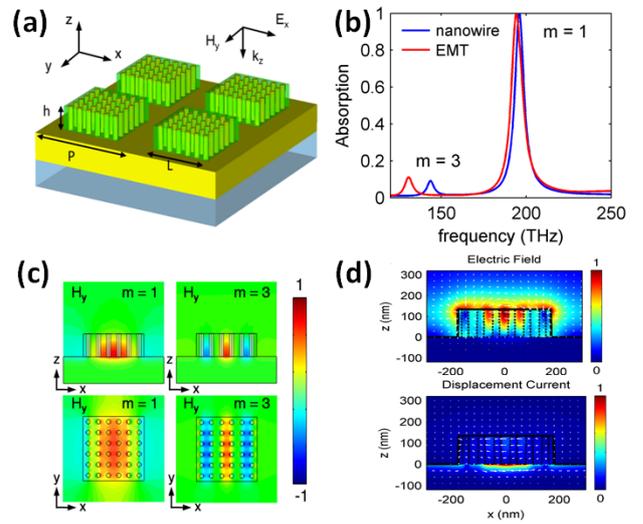

Figure 6 (a) Configuration of the metamaterial absorber comprised of cavities formed by a metallic nanowire array placed on top of a metallic substrate [152]. (b) Absorption spectra at normal incidence for the realistic absorber and the effective medium absorber, respectively. (c) Magnetic field profiles of the $m = 1$ and $m = 3$ resonant modes. (d) Distributions of the electric field (top panel) and the displacement current (bottom panel) when perfect absorption occurs.

The mechanism for perfect absorption is revealed by examining the mode profile. The magnetic field profiles $H_y$ in the x-z plane of the cavity resonant modes are shown in Fig. 6c, where the $m = 1$ mode with one magnetic field peak along the $x$ direction is located at the perfect absorption frequency; meanwhile the $m = 3$ mode corresponds to the weak absorption resonance. In the x-y plane, the magnetic field profiles are homogeneous along the $y$ direction for both modes. The $m = 2$ mode is absent at the normal incidence due to the structural mirror symmetry (with respect to y-z plane) of the metamaterial cavities.

The mechanism of perfect light absorption for the $m = 1$ mode can be understood as the cavity mode supporting the electric dipole resonance and the magnetic dipole resonance simultaneously. On one hand, the divergence and convergence of the electric field at the top left and top right corners of the metamaterial cavity (as shown in the top panel of Fig. 6d) implies the accumulation of positively polarized charges and negatively polarized charges, which gives rise to strong electric dipoles. On the other hand, strong anti-parallel currents are observed inside the metamaterial cavity and the grounded gold film, which leads to a strong magnetic dipole (as shown in the bottom panel of Fig. 6d). Since both the electric dipole and magnetic dipole are strongly excited, the effective impedance of a metamaterial absorber can be matched to that of the free space. By virtue of the strong resonances, the total absorption of incident light is achieved within a subwavelength thickness.

### 2.3.3 Ultrathin absorbers based on MNZ metamaterials

ENZ/MNZ ($\varepsilon$-near-zero/$\mu$-near-zero) metamaterials, of which Real($\varepsilon$)/Real($\mu$) approaches zero, have attracted much attention due to many abnormal phenomena. It was predicted that perfect absorption could be achieved by some arbitrarily thin ENZs and MNZs with low loss [154]. Another work shows perfect absorption in ultra-thin anisotropic ENZ metamaterials due to coherent perfect absorption via critical coupling to a fast wave propagating along the ENZ layer [155]. Recently, we have shown that an MNZ metamaterial with a large imaginary part could be superior in constructing absorbers as perfect absorption can occur at normal incidence [156]. Moreover, the sample has an electrical thickness of only about $\lambda/90$ and is 98% lighter than traditional microwave absorbers made of natural materials working at the same frequencies. In this subsection, the principle of this ultrathin MNZ metamaterial absorber is briefly presented.

Fig. 7a illustrates the corresponding ultrathin structure for achieving perfect absorption. An MNZ metamaterial slab with thickness $d$ is sandwiched between a semi-infinite dielectric layer and the bottom perfect electric conductor. The top layer and the slab are denoted as regions 0 and 1, respectively. For simplicity, assume region 0 to be the free space with permittivity $\varepsilon_0$ and permeability $\mu_0$. The relative permittivity and permeability tensors of the metamaterial in region 1 can be described by ($\varepsilon_{1x}$, $\varepsilon_{1y}$, $\varepsilon_{1z}$) and ($\mu_{1x}$, $\mu_{1y}$, $\mu_{1z}$), respectively. A TM-polarized plane wave impinges on the metamaterial slab with incident angle $\theta$. At normal incidence, the reflection can be expressed as

$$r = (1 + ik_0\mu_{1y}d)/(1 - ik_0\mu_{1y}d) \qquad (13)$$

Then, the solution for the perfect absorption condition is obtained by $\mu_{1y}=i\lambda/2\pi d$, where $\lambda$ is the wavelength in free space. The analytical solutions reveal that a metamaterial layer with a large purely imaginary permeability backed by a metallic plane has zero reflection at normal incidence when the thickness is ultrathin. Fig. 7b plots the angular reflections for different values of Real($\mu_{1y}$) [with Imag($\mu_{1y}$) fixed to 14.3]. It is found that any deviation of Real($\mu_{1y}$) from zero would cause higher reflection (i.e. lower absorption) for all incident angles. Thus, it is critical to maintain a vanishing real part of $\mu_{1y}$ to achieve large absorption.

Guided by the above analytical formula, an ultrathin metamaterial absorber using an MNZ material was designed and fabricated. Double-layered spiral shaped rings function as the unit cell of the MNZ metamaterials (Fig. 7c). The extracted effective permeability near the resonance frequency 1.745 GHz is 8.6$i$ and the electrical thickness $d$ is nearly $\lambda/90$ (Fig. 7d). The metamaterial slab is fabricated by using standard PCB photolithography, and then the slab is cut into tens of strips, followed by the strips being packed up onto a PCB with a 17 μm copper sheet as shown in Fig. 7d. The photograph is shown in Fig. 7e. The absorption performance of the tested sample in Fig. 7f is verified by measuring the complex $S$-parameters in a microwave anechoic chamber. As depicted in Fig. 7g, reflection dips are lower than -12 dB at 1.74 GHz for all

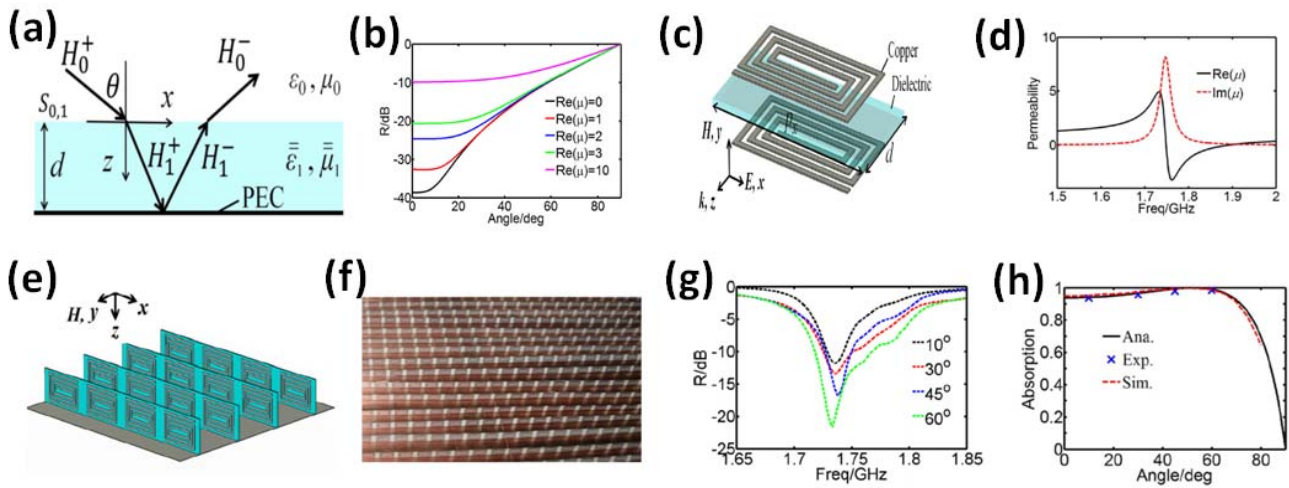

Figure 7 (a) Configuration of the theoretical model [143]. (b) Angular reflection spectra of absorbers with different Real($\mu_{1y}$). (c) Schematic of the dual-layer square spiral metamaterial. (d) The effective permeability of the unit cell. (e) Configuration for one-polarization absorption. (f) Photograph of the fabricated sample. (g) Experimental reflection spectra for the sample at different incidence angles. (h) Analytical, simulated and experimental angular absorptions at 1.74GHz.

incidence angles. Then, the experimental absorptions at 1.74 GHz are compared with the analytical and simulation results in Fig. 7h. Absorption of over 93% at incidence angles up to 60° has been achieved. The experimental results agree well with both the analytical and simulation results.

It is emphasized that the benefit of ultrathin MNZ metamaterial absorbers, compared with traditional absorbers made of natural materials, can bring great potential in many military and civil applications, such as stealth, camouflage and reduction of electromagnetic interference.

## 3 Broadband/multiband EM absorbers

Absorbers with excellent absorption within a narrow wavelength range can be easily obtained, but they reflect a large amount of incident light within a broad wavelength range. Such a property of wavelength sensitivity severely limits the application of EM absorbers in the areas of solar energy harvesting. To overcome this limitation, different methods have been proposed such as mixing multiple resonators within the same unit cell [24,48,131,161], exciting phase resonances to divide the absorption band into multiple sub-bands [95,162], employing tapered anisotropic metamaterial waveguides to stop light [25,27,163], using high-loss materials as the absorbing constituents [33,37,41,164,165], and others [23,159,166].

### 3.1 Mixing multiple resonances together

The simplest, as well as the most direct method, to expand the absorption band is to embrace multiple resonances working at distinct frequencies within the same unit cell, based on the concept of the collective effect of multiple distinct oscillators. The resonators mixed within each unit cell can work on the same principle but at different sizes [28,48,131,167-174]. Because the resonators combined together should have negligible influences on each other, their principles of absorbing light are on the base of the excitation of localized resonances, especially the gap SPP resonance excited within MIM cavities [31].

### 3.1.1 Mixing resonances by horizontal integration

Most works integrating multi-sized resonators within the same unit cell are in the form of parallel arrangement in the same plane [48,167-171,174-177]. Such an idea came into being since the absorption cross section of this type of resonance is large enough. Hence, the efficiency and bandwidth of each absorption peak remain very well after adding additional resonators in its vicinity. In particular, if the resonators are composed of hollow rings, they can be strung together for achieving efficient multiband absorbers [28,172,173,178]. Here, we note that when significant differences exist in size between neighboring resonators, one can only obtain multiband absorbers [28,167,169,172-174]. Otherwise, if the feature sizes of the constituent resonators have relatively small differences between each other, several adjacent narrow absorption bands can mix into one broad absorption band [48,168,170,171,175-177].

The following shows an example of the idea of horizontal integration. Fig. 8a shows the 3D configuration of a broadband light absorber comprised of an array of multi-sized gold nanostrips on top of a substrate evaporated with a reflective gold film and a 340 nm thick germanium film [168]. The gold nanostrips are of four different widths (labelled $W_{1-4}$) and patterned within one regular period. $W_1$ to $W_4$ form an arithmetic sequence, and the spaces between neighboring strips were kept identical. The superiority of this idea can be clearly displayed by studying a reference with a single-sized nanostrip (only the nanostrip of width $W_3$ is reserved). Fig. 8b plots the calculated absorption spectra under TM-polarized incidence of the reference and our multi-sized strip absorber, respectively. Evidently, one sees that the absorbing spectrum of the multi-sized strip absorber is very broad as a result of four close peaks (I, II, III and IV) merging together, of which peak III is located very closely to the resonant peak of the single-sized strip absorber. By calculation, it is obtained that the multi-sized strip absorber has an FWHM equal to 31.3% at the centered wavelength $\lambda = 9.84$ μm, while for the single-sized strip absorber, FWHM is only 9.2%, less than one third of that

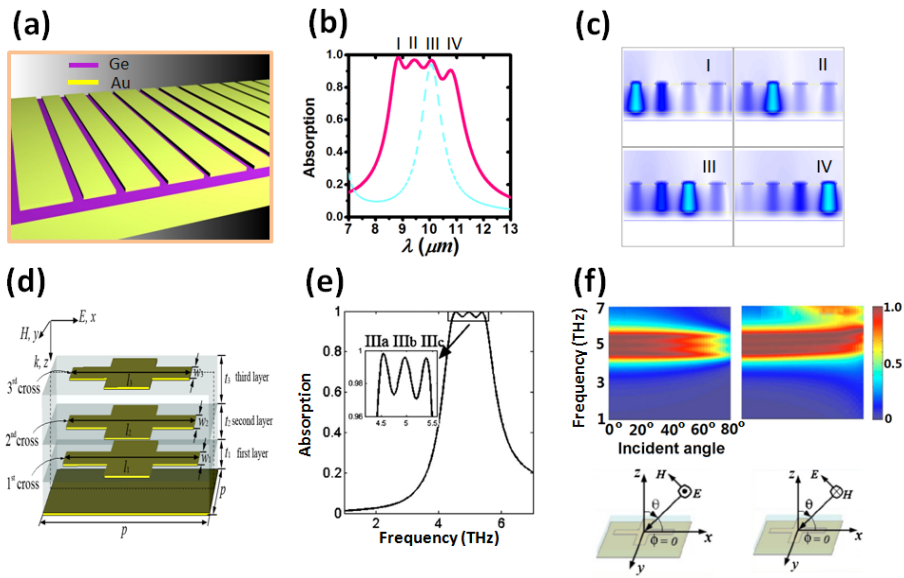

Figure 8 (a) Configuration of the horizontally integrated multi-sized MIM cavities [154]. (b) Absorption spectra at normal incidence for the multi-sized strip and single-sized strip absorbers. (c) Field distributions at the four absorption peaks in (b). (d) Configuration of the vertically integrated multi-sized MIM cavities [119]. (e) Absorption spectra at normal incidence for the vertically integrated absorber. (f) Angular absorption spectra at TE or TM polarizations for the vertically integrated absorber.

of the broad-banded one. Here, Fig. 8c shows the field distributions at the four peaks of the multi-sized strip absorber, respectively. It is clearly observed that the energy centre shifts from the region under the first strip to the fourth when the wavelength is tuned from peak I to peak IV.

Intuitively, the absorption bandwidth widens as the number of distinctly sized resonators increases. However, when high absorption efficiencies within the broadband wavelength range are ensured, the number of resonators combined within the same unit cell is limited since the absorption cross section of the resonators is finite. Thereby, the broadening effect based on this type of integration is restricted with a finite bandwidth. In Ref. [171], resonators of seven different sizes have been packed together, for which part of the absorption peaks are already lower than 80%.

### 3.1.2 Mixing resonances by vertical integration

Another way of expanding the absorption band is by vertically stacking different-sized resonators together within the same unit cell size, as we have demonstrated in [131]. Comparatively, vertical integration of the resonators does not have a limitation on the number of integrated resonators. Hence, the absorption bandwidth of the constructed absorber can be much broader.

In Ref. [131], the resonant units forming the broadband absorber are metallic crosses for the purpose of eliminating the polarization sensitivity. The crosses of slightly different lengths are stacked layer by layer to ensure that the resonance could be situated near the same frequency. Fig. 8d shows the schematic diagram of a 3-layer cross structure as an example, which consists of three alternating layers: gold cross, polymer spacer, and a gold film at the bottom. The wire lengths of the three crosses are equal to 17, 15.4, and 15 μm from bottom to top, respectively. The wire widths of all the crosses are always equal and fixed to 6 μm, for the consideration of simplicity. Then, by tuning the polymer spacer thickness of each layer, the multi-layer structure can be impedance-matched to the free space at each resonant frequency. The optimized thicknesses of the three polymer spacers are equal to 0.7, 1.1, and 2.0 μm from bottom to top, respectively. The corresponding absorption spectra for a normally incident plane wave with polarization along the *x* direction are simulated and presented in Fig. 8e. It is seen that there are three closely positioned resonances (III$_a$, III$_b$, and III$_c$), each with absorption up to 99.9%. To understand the origin of the spectral characteristics, the distributions of the magnitude of the magnetic field's *y*-component |$H_y$| in the $y = 0$ plane at the three resonances are investigated. It is found that each resonance is a hybridized mode, which is formed differently by three magnetic polaritons for each layer. For example, for the peak of III$_a$, the incident wave is strongly concentrated under the first cross while the fields under the second and third crosses are fairly weak. Significantly, the wide-angle feature of this high absorption caused by localized resonances is preserved in this multi-layer structure as shown in Fig. 8f for TE and TM waves, respectively. Meanwhile, the whole structure is designed with a four-fold rotational symmetry, which makes the absorption variation quite insensitive to the change in the azimuthal angle. A further broadening of the absorption bandwidth is also possible by increasing the number of stacked layers. However, there is a trade-off between the thickness of the absorber and the bandwidth of the absorption, and absorbers with more layers require higher fabrication cost.

### 3.1.3 Other methods to mix resonances

An interesting idea that does not include more than one resonator in each unit cell but also produces two resonant absorption peaks is by breaking the symmetry of the resonators [12,24,179,180]. Rectangular, elliptical, or asymmetrical cross-shaped structures with special arrangement have been adopted for designing dual-band absorbers. In fact, at a specific polarized incidence, its mechanism of absorption is similar to that of the aforementioned horizontal integration of resonators of two different sizes (of which the drawback is that it can only produce dual absorption bands). In addition, multiband absorbers can also be obtained based on resonances of different principles as well as the mutual coupling effect between different resonances [161,181-184]. For example, by placing a metallic nanostrip array above a metallic nanogroove array with a separation of 120 nm, one can obtain three different types of plasmonic resonances at different wavelengths, all of which display absorptivity greater than 90% [161]. A numerical study indicates that the multiple absorption peaks correspond to different types of resonances excited by the sole metallic nanostrip array, the sole metallic nanogroove array, and the coupling between the nanostrip and nanogroove, respectively. Another work to achieve polarization insensitive broadband absorbers based on mixing resonances was by carving the top thin metal film of an MIM multilayer system into the format of the so-called crossed trapezoid array [185]. Recently, Chen *et al.* developed an inexpensive droplet evaporation method to create patterns of gold nanorods dispersed on a dielectric loaded gold film [186]. By such method, they demonstrated an extremely broadband absorber for the 900–1600 nm wavelength range. The widen absorption was ascribed to the excitation of different plasmon modes generated from the random and sophisticated distribution of the multilayer of gold nanorods.

In summary, subsection 3.1 has introduced the recent progresses on overcoming the limitations of narrow absorption bands by mixing multiple resonances within the same unit cell. We have overviewed several different approaches to combine multiple resonances together, such as horizontal integration, vertical integration, broken symmetry, and multi-resonance combination. Among them, the approach of horizontal integration is relatively cost-effective, but its ability of broadening the absorption band is limited, while the vertical integration method can produce a much broader absorption band but the related processing techniques are complicated.

## 3.2 Exciting phase resonances based on lattice scattering

This subsection focuses on a novel type of resonances generated due to the effect of lattice scattering [187,188], which is advantageous in designing multiband absorbers. A single absorption band can split into multiple sub-bands, and the specific features of distinct resonant peaks are their unique phase resonances. We have reported two types of multiband absorbers resorting to the lattice scattering effect. One is an extension work on the base of the metal/dielectric stack absorbers presented in subsection 2.1, and another is the modification of conventional metallic surfaces of V-shape grooves as mentioned in subsection 2.2.2.

### 3.2.1 Metal-dielectric photonic crystal based absorbers

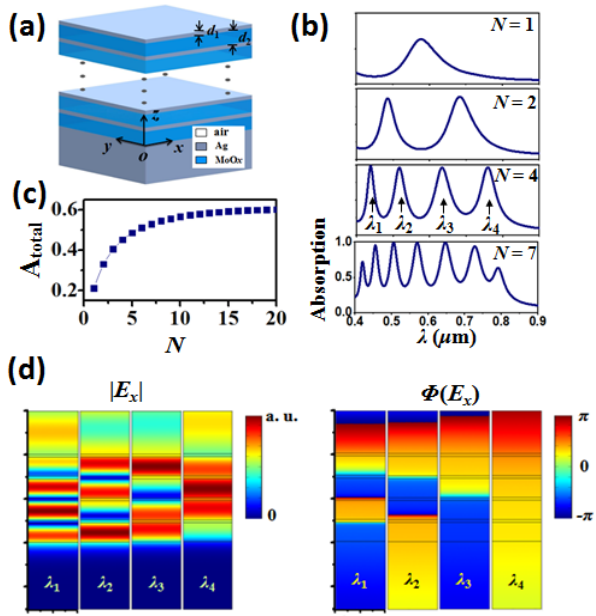

Figure 9 (a) Configuration of the planar absorber based on photonic crystals [149]. (b) Absorption spectra at normal incidence for the proposed absorber when the number of the unit cells ($N$) varies. (c) The total absorption efficiency ($A_{total}$) with varying $N$. (d) Distributions of the electric field $|E_x|$ and phase $\Phi(E_x)$ at the four absorption peaks of the 4-unit cell planar absorber.

In Ref. [56], the planar absorber comprises of only three layers, including one thin metallic film and one dielectric film on top of a thick metallic substrate. The modified proposal to excite multiple absorption bands is to place a periodic metal-dielectric photonic crystal (MDPhC) on top of the reflective substrate as shown in Fig. 8a. The absorption property of MDPhCs has already been discussed [189,190], and their simulated efficiencies at the absorption peaks are only around 50% due to the absence of the reflective substrate.

We have proposed a modified device with an additional reflective substrate recently [162]. In Fig. 9a, alternating Ag and $MoO_x$ slabs form the truncated MDPhC. The number of the unit cells is defined as $N$. Fig. 9b shows the absorption spectra at normal incidence when the number of unit cells varies from 1 to 8. As the value of $N$ increases, the number of peaks within the investigated wavelength range increases because there are more unit cells involved in the hybridization process. These absorption peaks lie in the first photonic band, and the resonances exist because the photonic crystal lattices have the function of selecting the resonant modes. For the selected modes, the component of the wave vector along the $z$ axis is approximately $j\pi/L$, in which $j$ is an arbitrary integer and $L$ is the total thickness of the MDPhC equal to $N(d_1 + d_2)$.

Take the case with $N = 4$ for further illustration. As shown in Fig. 9b, it has four absorption peaks, each of which has an efficiency higher than 95% and they (from longest to shortest wavelengths) correspond to the resonant modes with $j = 1, 2, 3,$ and 4, respectively. The amplitude along with the phase distributions of the $E_x$ field are plotted in Fig. 9d and it is found that there are four different types of phase profiles. A sign of '+' represents the phase close to $\pi/2$ and the sign of '-' the phase close to $-\pi/2$. The phase resonance for the peak at $\lambda_4$ can be indicated by '+', and is the fundamental order of resonant mode. Those of higher orders are excited at $\lambda_3, \lambda_2,$ and $\lambda_1$, being represented respectively by signs of '+-', '+-+', and "+-+-". Amplitude plots also shows that the number of the field nodes with a minimum field amplitude increases from 0 to 3 as the resonant wavelength shifts to the shorter wavelength range.

The total energy absorption gradually increases with more unit cells, but the rate of increase rate tends to reduce, as observed in the inset of Fig. 9c. Note that despite that the total absorption efficiency of the device with more unit cells is higher, it requires more materials and suffers higher fabrication cost. One way to overcome the limited absorption is by tuning the thicknesses of the silver layers. For an optimized device for the 8-unit cell MDPhC with gradually tuned silver slab thicknesses, all eight peaks have efficiencies higher than 90%. In addition, some reports demonstrated that the form of quasi-periodic MDPhCs [191] or aperiodic metal/dielectric stacks [35] can also produce absorption performance better than the MDPhC device without a reflective substrate. Another way to improve the broadband absorption performance is by employing anisotropic metamaterials based on high loss metals, which has been discussed in Ref. [157]. These kinds of planar absorption devices have promising futures in the area of thermal photovoltaics, requiring more investment in research and development.

### 3.2.2 Hybridization effect between neighbouring grooves

Lattice scattering can also be introduced in the metallic groove geometry, so that a single absorption band can split into multiple absorption bands [95]. As presented in subsection 2.2.2, metallic grooves with optically large depth can produce total absorption of light due to the excitation of localized SPPs. The shapes of the grooves are versatile. Here, V-shaped silver grooves composed of a wide groove (of width $W_1$ and height $h_1$) over a narrow groove (of width $W_2$ and height $h_2$) with periodicity $P$ are adopted to form the original metallic surface as shown in Fig. 10a. Such a structure produces only one absorption peak at the frequency of 0.42 eV within the range of [0.3, 0.6] eV at the normal incidence of TM-polarized light, as shown by the dotted line in Fig. 10c. In Fig. 10d, the left subplot shows the distributions of field magnitude $|H_y|$ and phase $\Phi(H_y)$ at its absorption peak, which shows that each V-groove supports the first order F-P resonance.

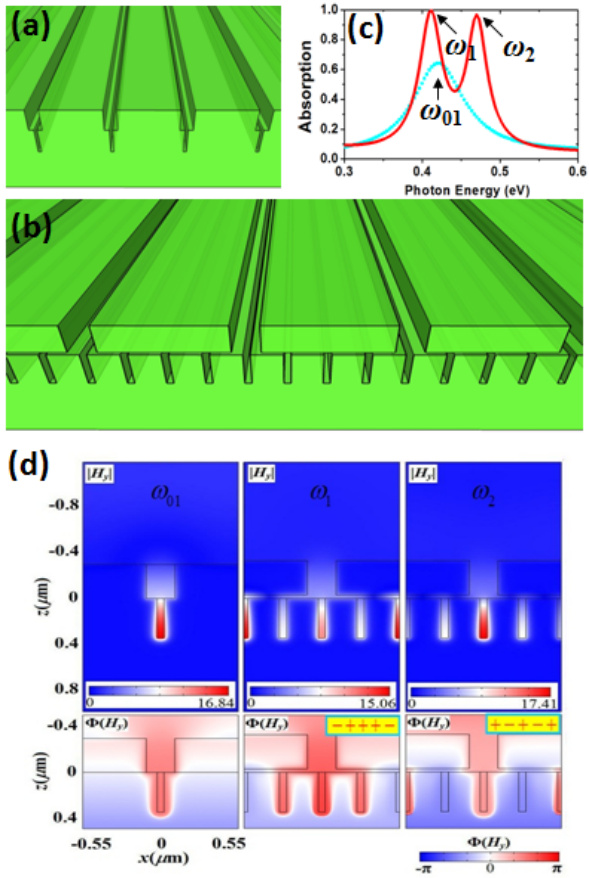

Figure 10 (a) Sketches of the original V-shaped groove metallic surface and (b) the modified absorber-supporting phase resonances [95]. (c) Absorption spectra as a function of photon energies for the original surface (dashed) and the phase absorber (solid). All of their absorption peaks are labelled. (d) Distributions of magnetic field $|H_y|$ and phase $\Phi(H_y)$ for the absorption peaks of $\omega_{01}$, $\omega_1$, $\omega_2$. Insets of $\Phi(H_y)$ in (d) show the phase states of each resonance near the opening of the bottom narrow grooves.

The modified structure is formed by cutting the metal horizontally at the bottleneck of the V-groove and drilling three additional narrow grooves, as shown in Fig. 10b. Such a structure has a narrow horizontal gap of width $d$. The additional narrow grooves are identical to the bottom part of the original V-shaped groove, and all the narrow grooves are equally separated. The corresponding absorption spectrum for the modified structure is shown by the solid line in Fig. 10c. It is observed that by introducing additional sub-lattices within each unit cell, the single absorption band successfully splits into two sub-bands located at 0.41 eV ($\omega_1$) and 0.47 eV ($\omega_2$), respectively. Both of the two peaks have much higher theoretical absorption efficiencies, greater than 96%.

The corresponding field distributions at the two absorption peaks for the modified device are shown in the middle and right subplots in Fig. 10d. Considering only the field distribution within the middle groove and the top part of the V-shaped grooves, one sees that the magnitude and phase distributions at $\omega_1$ and $\omega_2$ are very close to that of the original case. Therefore, for the modified metallic surface, although a horizontal gap geometrically breaks the original T-groove into two parts, this gap of $d = 30$ nm is too small to alter the field distribution within the V-shaped groove. In other words, the resonances at $\omega_1$ and $\omega_2$ are all generated from resonance $\omega_{01}$. However, these two resonances are different in terms of the phase information along the $x$-direction at the openings of the narrow grooves. If we consider the phase to be near the opening of each groove, the phase distribution at $\omega_1$ and $\omega_2$ can be represented by '- + + + -' and '+ - + - +', respectively (the sequence of signs corresponds to the field sign at the opening of grooves #1, #2, #3, #4, and the next #1 at a certain time). It is found that within each period there is a phase change of   and $2\pi$ along the $x$ direction at $\omega_1$ and $\omega_2$, respectively. This suggests that their resonances have wavenumbers 2 /$P$ and 4 /$P$, respectively. The excitation of finite parallel momentum at normal incidence is exactly due to the lattice scattering between neighboring, closely placed narrow grooves.

### 3.3 Employing anisotropic metamaterials to stop light

One of the most efficient methods of broadening the absorption band is by employing the array of tapered lossy metamaterial waveguides to "stop" light [25], (actually light of a specific wavelength is slowly "squeezed" and consequently absorbed effectively at a certain position of a tapered lossy metamaterial waveguide before it is reflected [192,193]). Fig. 11a shows the configuration consisting of alternating layers of flat metal and dielectric plates, which are etched into a sawtooth shape with the tooth widths increasing gradually from top to bottom. Through the excitation of the slow light [192-194] waveguide modes at different positions of the AMM sawteeth, the obtained absorption band with efficiency near unity is ultra-broad and quite flat.

This is actually a follow-up derivation from the broadband absorbers based on the three-layer-stacked resonances mentioned in Section 3.1.2 [131]. However, the principle of broadband absorbers is no longer the direct superposition of a series of narrow absorption bands, because the thickness of each dielectric slab is too thin to support any localized resonances. Here, metal plates are made of gold with a thickness of 15 nm; dielectric plates are made of germanium with a thickness of 35 nm. The total number of metal/dielectric pairs (N) is 20. A gold film with an adequately large thickness is added under the sawtooth anisotropic metamaterial (AMM) slab to block all transmission. The absorption spectrum at normal incidence for TM-polarized light (Fig. 11b, thick line) indicates that the absorption performance is excellent, with absorptivity higher than 95% covering the range from 3 to 5.5 μm. Its full absorption width at half maximum (FWHM) is around 86%, five times that of a typical narrow band absorber based on metallic crosses (only 16.7%) [29]. Because each of the composition layers is very thin, the metal/dielectric multilayer system can be described by the parallel and the perpendicular effective permittivities. Then, the sawtooth device with effective parameters is constructed, and it is found that its absorption spectrum (Fig. 11b, thin line) is very similar to that of the actual sawtooth structure. In addition, the performance of very high and ultra-broadband absorption retains very well even when the incident angle is 60° as plotted in Fig. 11c.

For the purpose of understanding how light is absorbed in the sawtooth AMM absorber, the normalized field distributions $|H_y|$ in the actual inhomogeneous sawtooth structure are shown by the color maps in Fig. 11e with $\lambda_0 = 3.5$, 4.5, and 5.5 μm, respectively. It is identified that light of different wavelengths accumulate (actually are squeezed) at the parts of the sawtooth absorbers of different waveguide widths. Plots of Poynting vector (**S**) (arrows in Fig. 11e) reflect that most of the incident energy first propagates downwardly along the z direction in the air gap region without penetrating into the AMM sawteeth and then whirls into the AMM region, forming vortices close to the interface between the AMM and air regions. The vortices are located exactly at the places where the magnetic field is concentrated. These are typical features in slow-light waveguides.

The physical principle of ultra-broadband absorption can also be understood based on with the following analytical theory as well. For a three-layered air/AMM/air waveguide with a core width of W, its dispersion relationship between the incident photon frequency ($\omega_c = \omega/c$) and the propagating constant ($\beta$) is derived as

$$\exp[iq_2 W] + \frac{\frac{\kappa_1}{n_0^2} - i\frac{q_2}{\varepsilon_\perp}}{\frac{\kappa_1}{n_0^2} + i\frac{q_2}{\varepsilon_\perp}} = 0 \quad (14)$$

where $\kappa_1 = \sqrt{\beta^2 - n_0^2 \omega_c^2}$, $q_2 = \sqrt{\varepsilon_\perp \omega_c^2 - \frac{\varepsilon_\perp}{\varepsilon_\parallel}\beta^2}$, and $n_0$ is the refractive index of air. Dispersion diagrams of the fundamental order of the waveguides with different fixed core widths can be found in Fig. 11d. One observes that for each dispersion curve, at a lower frequency band the propagating constant increases gradually with the incident photon frequency; however, when the incident light approaches the cut-off, the dispersion curve becomes flat. A close-up view demonstrates that these dispersion curves are not strictly flat around the cut-off frequencies but decline slightly after passing them. This indicates that the fundamental propagating mode has an extreme point with the group velocity approaching 0 (i.e. the slow light modes). For a waveguide of a certain core width, the slow light mode (of zero group velocity) is generated around a certain photon frequency (or wavelength). Therefore, the

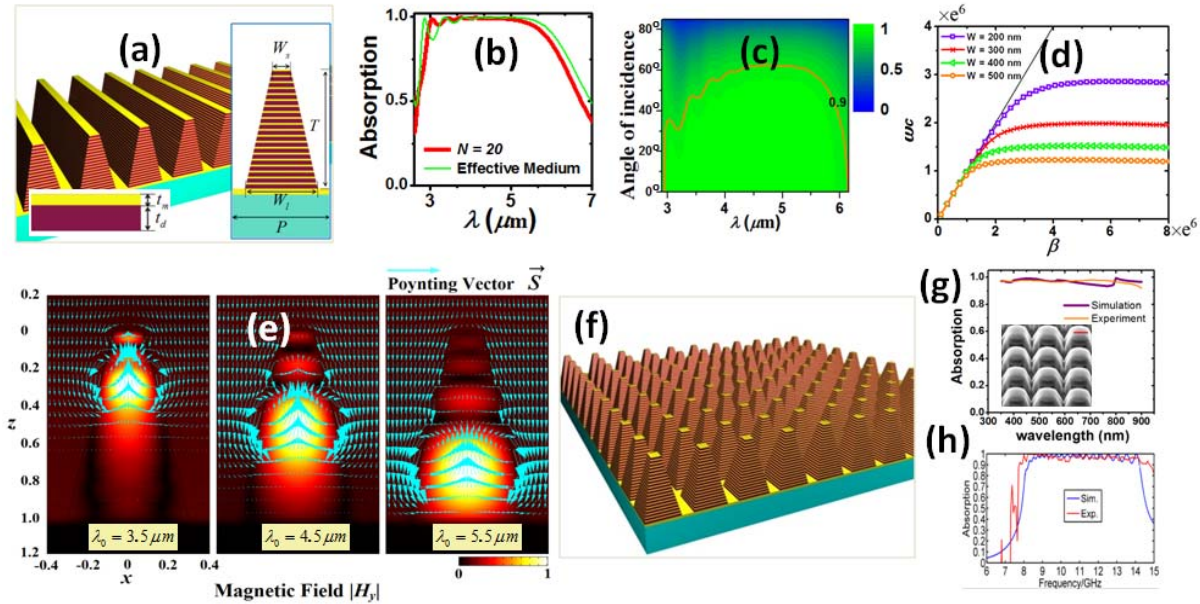

Figure 11 (a) Diagram of the sawtooth anisotropic metamaterial absorber [25]. (b) Absorption spectra for the sawtooth AMM absorber with number of periods $N = 20$ (thick) and the effective homogeneous sawtooth structure (thin). (c) Angular absorption spectrum with the line representing the efficiency contour with $\eta = 0.9$. (d) Dispersion curves of the air/AMM/air waveguides at different $W$. (e) Distributions of magnetic field (color) and energy flow (arrows) for the sawtooth AMM absorber at different incident wavelengths. (f) 3D diagram of the absorber made of a 2D array of an AMM pyramid. (g) Absorption spectra for a Vis-NIR absorber made of a 2D array of an AMM pyramid. The inset shows its SEM picture (scale bar 300 nm). (h) Comparison between the experimental absorption (red) and simulated absorption (blue) for a microwave absorber based on AMM pyramids.

AMM sawtooth absorber with a tapered width can be regarded as a group of air/AMM/air waveguide absorbers of infinite number, which can respond at different frequencies and then display as a collective effect.

When the schematic diagram is extended into a 2D array of quadrangular frustum pyramids, its dependence of absorption on polarization can be eliminated (Fig. 11f). By resorting to focused-ion-beam (FIB) technology, we have successfully demonstrated a Vis-NIR absorber made of a 2D array of AMM pyramids as shown in the inset of Fig 11g. An AMM comprising of 5 pairs of $Al_2O_3$ and Cr layers is deposited on top of a gold mirror. The fabricated geometries are $W_s$ = 300 nm, $W_l$ = 750 nm, $P$ = 800 nm, $t_m$ = 10 nm, and $t_d$ = 70 nm. It is impressive to find that the measured absorption (Fig. 11g, thin line) is above 90% in the spectrum range from 350 to 900 nm, in good agreement with the simulation results (Fig. 11g, thick line). We have also successfully demonstrated the same concept in the microwave regime [27]. An example was fabricated using the standard PCB technology, which has an area of 200 mm × 200 mm, including 256 units. The parameters are $W_s$ = 5 mm, $W_l$ = 9 mm, $P$ = 11 mm, $t_m$ = 0.05 mm, $t_d$ = 0.2 mm, and $T$ = 5 mm. The absorption performance of the test sample was verified by measuring the complex S-parameters in a microwave anechoic chamber with the experimental absorption spectrum shown in Fig. 11h by the red curve. One can see good agreement between the experimental and simulated (blue curve) results considering the tolerance in the fabrication and assembly. By exciting higher order resonances, even a blackbody working from ultraviolet to mid-infrared, covering the visible range, is also achieved [195]. It is expected that the design based on AMMs would find great applications in practice due to its remarkable property of extending the working bandwidth in the process of EM wave absorption.

### 3.4 Utilizing high loss materials as the absorbing constituents

Since the objective is to achieve good absorption, intuitively, using high loss materials should be advantageous for producing the desired effect than using noble metal alone. For example, adopting structured high loss metals, e.g., tungsten, nickel, and tantalum, can generate very good absorption performance [37,157,164,196-198]. Besides, there already exist a series of reports utilizing low-bandgap semiconductors (e.g. silicon and germanium) to form good broadband absorbers. Generally, semiconductors are etched into some specific micro-/nano-structures, such as wires, holes, tips, cones, etc., to bring forth the antireflection effect and further improved absorption properties [33,165,199-202]. In the following, we will refer to two specific instances of broadband absorbers based on high loss materials.

One effort to achieve broadband absorbers is by replacing part of the components made of noble metals (e.g. Au or Ag) by counterparts made of high loss metals. For a typical plasmonic metamaterial absorber comprised of three metal/insulator/metal (MIM) layers, its absorption bandwidth is very narrow [31], as plotted by the thin line in Fig. 12b. Optimization is achieved by replacing the top patches made of gold by Ti-made patches as shown in Fig. 12a. Then, the geometrical parameters are modified accordingly. The thicknesses of the silica spacer and the gold film are 160 nm and 40 nm, respectively. The diameter and thickness of the titanium disks are 400 nm and 30 nm, respectively. The periods in both the $x$ and $y$-directions are 600 nm. For such a structure, the simulated absorption spectrum is shown by the thick line in Fig. 12b. It is shown that the absorptivity is above 90% from 0.9 to 1.8 μm, and the obtained absorption band width is much broader than the original device based on gold disks. Similar work has also been done using tungsten disks [164].

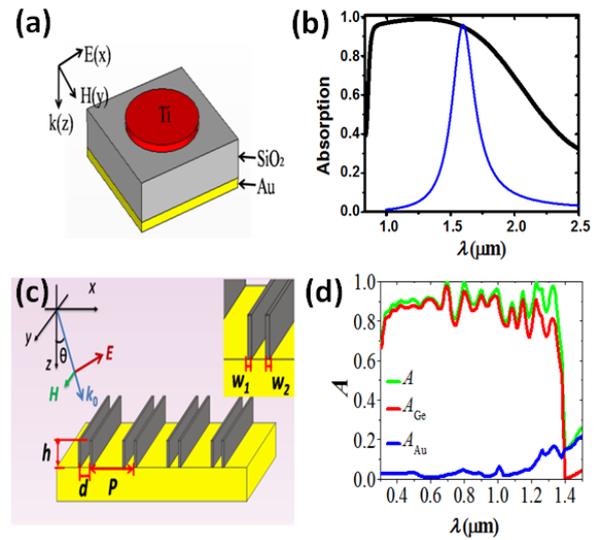

Figure 12 (a) Broadband absorber based on MIM cavities with the top layer of metal made of Ti. (b) Absorption spectra for the Ti-based absorber (thick). The thin line represents the corresponding results shown in Fig. 5b. (c) Schematic diagram of an absorber based on a periodic array of Ge nanowire pairs [203]. (d) Absorption spectra by the whole device, Ge and Au for the Ge-based absorber.

In the following, a Ge related broadband absorber is introduced. We put a Ge-made dielectric slot waveguide grating (SWG) on the metallic substrate [203]. In detail, the SWG consists of two germanium nanowires (Ge NWs) separated by a sub-100 nm slot in each period as shown schematically in Fig. 12c. As a reflector, the Au substrate is assumed to be thick enough to avoid any transmission through it. Due to the fairly small separation between the two NWs in each period, slot waveguide modes together with some interesting photonic responses can be generated. As shown in Fig. 10d, high absorption of approximately 90% can be obtained over a broad wavelength range from 0.3 to 1.4 μm for TM polarization under normal incidence. The absorption retains very well even when the incident angle is as large as 80°. From Fig. 12d, it is seen that the absorption of Ge dominates the total absorption in the wavelength range below 1.4 μm. Multiple optical phenomena, i.e., diffraction, waveguide modes in the high-index Ge NWs and low-index air slot, FP resonances and SPPs, are identified to govern the absorption characteristics of this absorber.

## 4 Applications of Plasmonic/metamaterial absorbers

Plasmonic and metamaterial absorbers have wide applications in many areas, such as solar energy harvesting, thermal emitters, organic light emitting diodes, light coupling, chemical sensing, THz modulators and so on.

### 4.1 Solar energy harvesting

One notable aspect of solar energy harvesting is in photovoltaic cells. In this area, designs of EM absorbers that mainly convert light into ohmic heat by metals are improper. In practice, people usually adopt the mechanisms of EM absorptions to improve the ability of solar-active layers to trap light. This is because some low-cost solar cells have very thin active layers [204], but there is a mismatch between the optical absorption length and the carrier diffusion length [205]. To compensate for such a mismatch in length, light trapping becomes critically important to improve the whole device performance both optically and electronically [38,39].

Different principles are involved in the process of enhancing absorption: extending the optical length by scattering, enhancing filed intensity by localized SPPs, or converting the direction of the optical path by exciting planar SPP modes. For instance, arrays of horizontal MIM cavities have been applied in improving the efficiency of thin solar cells by simply replacing the insulator with certain active materials [40,101]. In 2012, Salinas et al. introduced such an FP cavity into polymer solar cells [209], achieving an overall external quantum efficiency comparable to that with ITO. More importantly, the resistivity and mechanical ductility of semitransparent metal films are excellent and thus the device shows excellent bending stability. It has also been demonstrated that by periodically patterning the FP cavity in the polymer solar cells, the short circuit current as well as the power conversion efficiency can be greatly enhanced compared with the control device with a flat shape [210,211]. Besides, many designs have displayed that patterning the metal electrodes into gratings can also significantly boost light absorption by active layers [41,206-208].

In addition to solar cells, the incident solar light can also be converted into storable energy by the concentrated solar thermal system as well as the thermal photovoltaic system [35,37,157,164,196-198,212-215]. In these systems, some of the broadband metallic EM absorbers presented in this review can be widely applied after optimization for absorbing the incident solar light within an extremely broad wavelength range [35,157,197,198,213].

### 4.2 Thermal emitters

The concept of the planar metal/dielectric structure as shown in Section 2.1 was first put forward for making a selective thermal emitter in 2006. Later, some complicated geometrical profiles were also utilized to design emitters [49,216]. In a thermal photovoltaic system, a selective narrowband emitter is required to provide a sharp emissivity peak at the solar cell band-gap [157,214]. With this purpose, the mechanisms of narrow band absorbers can be used. It has been noted that under a suitable light concentration condition, and with a reasonable area ratio between the emitter and absorber, a designed thermal photovoltaic system can achieve an efficiency that exceeds the Shockley-Queisser limit [214]. In 2011, Liu et al. have designed broadband thermal emitters based on the concept of mixing resonances [48].

### 4.3 Extracting light from OLEDs

The application of EM absorbers in the area of OLEDs can be traced back to a decade ago. In 1996, people suggested using very thin metal layers to make semi-transparent anodes, considering their friendly fabrication process in comparison with their counterpart of transparent conducting oxides [217]. Due to the effect of the FP cavity formed by the semi-transparent metallic anode and the thick metallic cathode, the emission characteristics, such as the emissive wavelength, the viewing angle, and the luminescent spectra can be flexibly tuned [218-220]. In OLEDs, such effect is usually called the microcavity effect. Metallic grooves have also been employed to increase the extracting efficiency of OLEDs [210,211,221]. In detail, by making the FP cavity [217] into a wrinkled shape, an enhanced light transmission through a 45 nm thick Ag cathode is obtained ascribed to light coupling between the grating-induced SPP mode and the microcavity mode [221]. In practice, a thick Ag film has great advantages in resisting diffusion of water and oxygen from air into the device. However, it suffers the drawback of low transmission. This work is very meaningful because it improves the capability of light transmission through the thick Ag film.

### 4.4 Light coupling

Angular selective absorbers comprised of shallow metallic grooves can play the role of coupling the incident light into planar SPPs in relation to research on extraordinary optical transmission of light through subwavelength metallic apertures [222-224]. This leads to a large amount of energy being coupled into metallic apertures. Moreover, the same grating can also be introduced around the output aperture so that the radiation of light can be controlled [224-226]. Due to the reciprocity theorem [226], the far field radiating angle can be obtained from the angle at which the maximum absorption is excited when light impinges on the surface of the metallic grating. Dielectric grating on top of a flat metallic substrate has also been applied in coupling light in or out of the metallic apertures [80,227,228].

### 4.5 Refractive index sensing

EM absorbers, especially the angular selective absorbers with very narrow absorption bands, are good candidates of sensing components. It is worth mentioning that the propagating loss of waveguide modes is much less than that of planar SPPs. In Ref. [81], the imaginary part of the propagating constant for the 1$^{st}$ order TM waveguide mode can be minimized to near zero when the thickness of the dielectric waveguide is just above the cutoff thickness. As a result, they experimentally obtained an absorption peak with a bandwidth of approximately 0.12 nm. The quality factor of such a high-finesse resonance is predicted to go beyond $10^4$. We note that, compared with the configurations based on planar SPPs, which are usually applied in refractive index sensing [229,230], the configurations with the excitation of waveguide modes might have a large figure of merit for their narrow absorption bands. In addition, based on cylinder-shaped cavities buried in metal, a theoretical nanofluidic refractive-index sensor was proposed [231]. It is worth mentioning that the excitation of the Ralyleigh anomaly [96] can also produce very sharp absorption peaks, which might have great potential in the application of refractive index sensing.

**4.6 Surface enhanced Raman scattering**

In the areas of surface enhanced Raman scattering (SERS) or fluorescent spectroscopy, EM absorbers that can excite extremely strong electric fields in the local region are very favorable [122,182,232-234]. With such an application, localized SPPs with optical deep grooves, instead of planar SPPs, should be excited in the sensing process. For instance, it was found that V-shaped grooves, which focus light into their sharp corners, can greatly enhance the Raman scattering signals [232].

**4.7 Other applications**

Moreover, EM absorbers can be applied in high performance infrared sensors [51], biosensors [108], modulators [53], single photon emission [235], photonic detectors [236,237], etc. For example, a THz modulator has been proposed by controlling the carrier density in the n-doped semiconductor spacer between a top-layer with patterned metallic patches and a metallic ground plate [53]. It was found that the absorption varies greatly with different applied voltage biases. The amplitude of the reflected wave, instead of the transmitted wave [238], was taken as the modulated signal.

**5 Summary and outlook**

The increasing pace in the development of EM absorbers has already resulted in many important achievements. In the past decade, a wide variety of EM absorbers have been put forward based on metallic sub-wavelength structures, and the capability of light trapping by metallic structures is acknowledged to be excellent. Ascribed to the versatility of forms of plasmonic or photonic modes, the mechanisms of EM absorbers are diverse. Designs required in relation to different application scenarios are usually different. Fortunately, whether the band is required to be broad or narrow, angle insensitive or sensitive, all can be realized by EM absorbers with specific geometries. In this article, we have reviewed almost all of the mechanisms related to EM absorbers, and this contributes to further research as a useful guideline.

Many design works highlight the ideas behind the wave absorption with special properties as well as instructions on how to optimize the structural parameters for the desired performance. However, in reality, some absorbing devices proposed theoretically with excellent performances can hardly be demonstrated by experiment, limited by the processing techniques available today, especially for those working in the optical frequency. Furthermore, future industrialization requires that the metallic sub-wavelength structures can be achieved with high yield in large areas and at low cost. The most profound application area of EM absorbers is solar energy harvesting. Every improvement to the performance of solar energy harvesting devices is of great significance to society, bringing both economic and environmental benefits. Significant research efforts have already been made around the world to improve their performances. In the future, low cost, easily fabricated, and high performance solar absorbers will be in high demand for building economic solar plants. With this background, physicists have adopted a few chemical processing technologies (e.g. self-assembly techniques) to fabricate periodic or quasi-periodic dielectric patterns and then transform them into patterned metallic structures. In addition, preparation methods of the sol-gel technique have also been introduced in synthesizing light trapping components. Overall, while high performance EM absorbers have great potential in many applications, their industrial realization remains a formidable challenge to overcome.

**Acknowledgement**

This work is partially supported by the NSFC of China (11204205, 91233208, 61178062, 60990322, 61274056, and 11204202) and the National High Technology Research and Development Program (863 Program) of China (No. 2012AA030402). Cui also acknowledges the New Teachers' Fund for Doctor Stations (Ministry of Education) (20121402120017), Hong Kong Scholar Plan (XJ2013002), the Top Young Academic Leaders of Higher Learning Institutions of Shanxi, Natural Foundation of Shanxi (2012011020-4), and '131' Leading Talents in Shanxi Universities. Cui thanks Dr. Kin Hung Fung for helpful discussions.

Key words: absorbers, plasmonics, metamaterials, broadband, subwavelength, metallic structures, thin film